\documentclass [amssymb,amsmath, showpacs]{revtex4-1} 
\usepackage{graphicx,epsfig,amsfonts,amssymb}
\usepackage{bm}
\usepackage{times}
\usepackage{lipsum}

\newcommand{\be}{\begin{equation}}
\newcommand{\ee}{\end{equation}}
\newcommand{\bes}{\begin{subequations}}
\newcommand{\ees}{\end{subequations}}
\newcommand{\bea}{\begin{eqnarray}}
\newcommand{\eea}{\end{eqnarray}}
\newcommand{\ba}{\begin{array}}
\newcommand{\ea}{\end{array}}
\newcommand{\beqn}{\begin{eqnarray*}}
\newcommand{\eeqn}{\end{eqnarray*}}

\newcommand{\la}{\langle}
\newcommand{\ra}{\rangle}

\newcommand{\dg}{\dagger}
\newcommand{\rar}{\rightarrow}

\begin{document}

\title{The Quest for Solvable Multistate Landau-Zener Models}

\author {Nikolai~A. {Sinitsyn}$^{a}$ and Vladimir~Y. Chernyak$^b$}
\address{$^a$ Theoretical Division, Los Alamos National Laboratory, Los Alamos, NM 87545,  USA}
\address{$^b$ Department of Chemistry and Department of Mathematics, Wayne State University, 5101 Cass Ave, Detroit, Michigan 48202, USA}

\begin{abstract}
Recently, integrability conditions (ICs) in mutistate Landau-Zener (MLZ) theory were proposed \cite{six-LZ}. They describe common properties of all known solved systems with linearly time-dependent  Hamiltonians.  Here we show that ICs enable efficient computer assisted search for new solvable MLZ models that span complexity range from several interacting states to mesoscopic systems with  many-body dynamics and combinatorially large phase space. This diversity suggests that nontrivial solvable MLZ models   are numerous. In addition, we refine the formulation of ICs and extend the class of solvable systems to models with points of multiple diabatic level crossing.
\end{abstract}
\date{\today}

\maketitle

\section{Introduction}
 



The future of quantum technology will  depend crucially on our ability to manipulate complex quantum systems by  time-dependent fields.  Achieving sufficient  control of driven quantum matter  is thus an important challenge, including to mathematical physics.
 For evolution of more than two interacting states, one of the difficulties here is the  scarcity of  solvable models that are described by the nonstationary Schr\"odinger equation with  practically interesting explicitly time-dependent Hamiltonians. Most of known solvable driven models are  either trivially reducible to independent harmonic oscillators and two-state systems   \cite{maj,multiparticle,reducible,yuzbashyan-LZ,galitski} or  correspond to almost impossible for implementation driving protocols such as multistate shortcuts to adiabaticity  \cite{shortcuts}. This lack of nontrivial useful results strongly restricts our understanding of quantum dynamics in explicitly time-dependent fields, especially at the level of  mesoscopic interacting systems with a combinatorially large accessible  phase space. 

Considerable progress on resolving this problem has been achieved within the MLZ theory, whose goal is  to find the scattering $N\times N$ matrix $\hat{S}$ that describes evolution with a linearly time-dependent Hamiltonian \cite{be}:
\begin{equation}
i\frac{d\Psi}{d t} = \hat{H}(t)\Psi, \quad \hat{H}(t) = \hat{A} +\hat{B}t .
\label{mlz}
\end{equation} 
Here, $\Psi$ is the state vector in the space of $N$ states; $\hat{A}$ and $\hat{B}$ are constant Hermitian $N\times N$ matrices.  
Linear time-dependence of parameters is relatively easy to implement in experiment, so MLZ models find numerous practical  applications  \cite{apps}.

One can always choose the, so-called, {\it diabatic basis} in which the matrix $\hat{B}$ is diagonal, and if any pair of its elements are degenerate then the corresponding off-diagonal element of the matrix $\hat{A}$ is zero.  
In the diabatic basis, elements of the matrix $\hat{B}$  are called the { slopes of diabatic levels},  nonzero off-diagonal elements of the matrix $\hat{A}$  are called the {coupling constants},  and diagonal elements of the Hamiltonian
are  called {diabatic energies}. 

An element $S_{nn'}$ of the scattering matrix is the amplitude of the state $n$ at $t  \rightarrow +\infty$, given that at $t \rightarrow -\infty$ the system was in the $n'$-th diabatic state. In many applications, only the matrix $\hat{P}$, with elements $P_{nn'}\equiv P_{n' \rar n} \equiv  |S_{nn'}|^2$  called  { transition probabilities}, is needed. More precise definition of the scattering matrix and discussion of its general properties can be found in previous reviews \cite{be,constraints}. Historically, the first study of  a model type (\ref{mlz}) beyond $N=2$ was done by Majorana in \cite{maj}. 

We will call a particular choice of Eq.~(\ref{mlz}) { solvable or integrable} if one can write a compact analytical expression for elements of the transition probability matrix in terms of the well understood special functions of model parameters.  Although the general solution of  MLZ problem remains unknown, the number of exactly solved realizations of Eq.~(\ref{mlz}) has been growing quickly recently due to the
discovery of {\it integrability conditions} (ICs) that identify models whose scattering matrices can be found exactly in the form of a product of specific elementary matrices \cite{six-LZ,four-LZ,cQED-LZ}. 
Unfortunately, there is still no known algorithm to derive systems that satisfy ICs.  Therefore the search for solvable 
models has required many trial-and-error efforts, so far.   

In this article, we describe tricks that we found useful to alleviate the latter problem. While there is still no algorithm for detailed classification of solvable MLZ models, our goal is to show that there are systematic ways to search for new models that satisfy ICs. For illustrations, we derive and investigate several new solvable cases with different patterns of energy level crossings and different sizes of the phase space. We suggest analytical expressions for elements of the transition probability matrices in these models and provide results of rigorous numerical tests of our solutions.

We would like to stress that ICs in MLZ theory have not been proved mathematically rigorously yet. However, computational algorithms, described in Ref.~\cite{cQED-LZ}, can be used to test our predictions with high precision. For example, transition probabilities in models of the type (\ref{mlz}) with up to $N=10$ states can be found numerically with up to three significant digits  precision within a few minutes using a standard PC. Numerical algorithms allow  ``high throughput" tests of our predictions for a broad range of model parameters and different initial conditions. We considered such a test as confirming the theory only when all tested multiple choices of parameters  produced numerical results that were indistinguishable from all our analytical predictions for a given model. We tested usually more than 100 different choices of constant parameters in the nonperturbative regime per model.
Thus, although this article is about mathematical questions, it is experimental in spirit. The role of an experimental setup is played here by a computer and results of our numerical experiments are obtained in order to provide an input for more mathematically rigorous studies of MLZ theory in the future. 

 The plan of our article is as follows. In section~\ref{s-prelim} we summarize basic knowledge about integrability in MLZ theory and suggest refined version of ICs. 
In section~\ref{s-four}, we demonstrate our approach using a simple four-state MLZ Hamiltonian.  
In section~\ref{s-six}, we show how one can generalize this approach to a higher dimensional case using a six-state model as an example, and in section~\ref{s-fermions} we show how similar intuition can be used to derive a new combinatorially complex but solvable MLZ model. We then summarize our findings in  Discussion.
 
 \section{MLZ models and integrability}
 \label{s-prelim}
 In this section, we review basic established facts about integrability in MLZ theory using two previously solved models as examples. We will frequently use this section for references later. The additional goal is to refine definition of  ICs.
 
 \subsection{Diabatic level diagram}
\begin{figure}
  \scalebox{0.35}[0.35]{\includegraphics{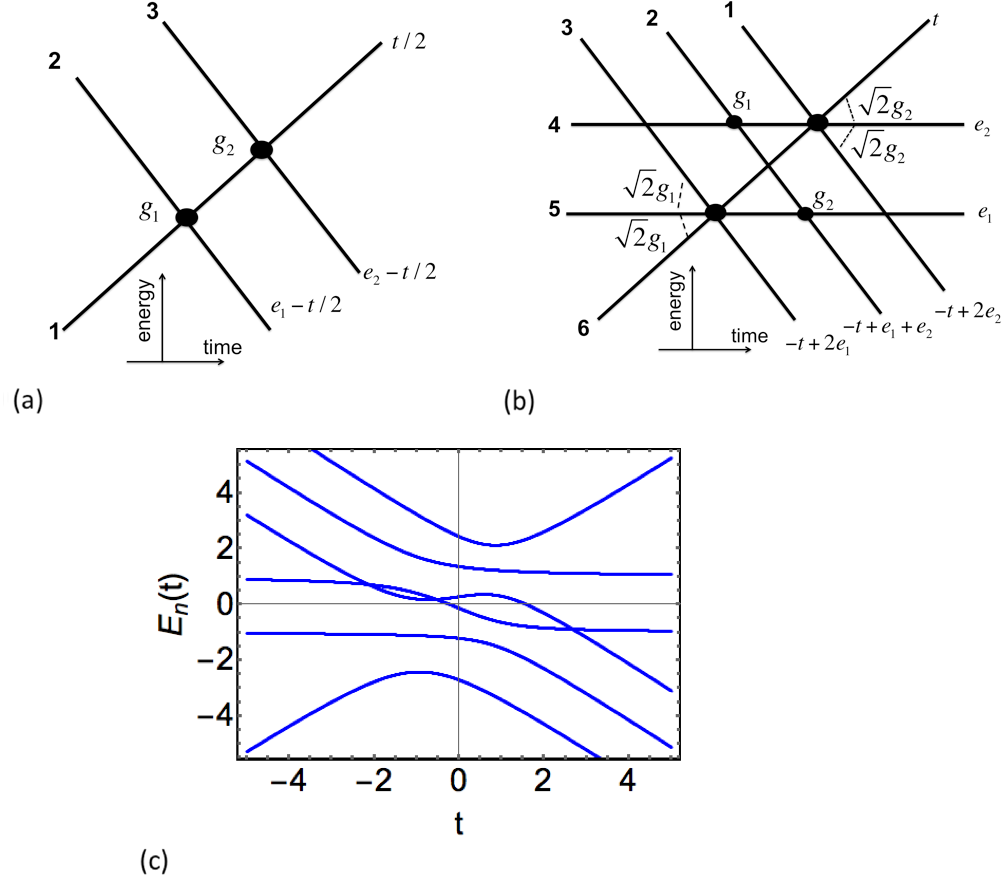}}
\hspace{-2mm}\vspace{-4mm}   
\caption{ (a-b) Diabatic level diagram for Hamiltonians in (a) Eq.~(\ref{h-do3}), and (b) Eq.~(\ref{do-b2-ham}). Straight lines are diagonal elements of the Hamiltonian as functions of time (diabatic levels). Pair-wise couplings between levels are marked at corresponding level intersections. Integers on the left mark level indexes and formulas on the right define energies of diabatic levels. (c) Adiabatic energy levels of the Hamiltonian (\ref{do-b2-ham}) as functions of $t$. Presence of three exact crossing points confirms validity of IC (ii). Parameters: $e_1=-1$, $e_2=1$, $g_1=0.67$, $g_2=0.7g$.}
\label{b1-fig}
\end{figure}

 It is convenient to illustrate parameters of any MLZ model in a graph with time-energy axes, as shown in Fig.~\ref{b1-fig}(a) and  Fig.~\ref{b1-fig}(b). 
Lines of this graph are time-dependent diabatic levels, i.e. diagonal elements of the Hamiltonian in the diabatic basis as functions of time. Small black filled circles mark intersections of levels with nonzero pair-vise couplings. Integers on the left side of diabatic levels mark level indexes. On the right, levels are sometimes 
marked by analytic expressions for diabatic energies. 
If intersection of two levels is not specially marked then corresponding coupling between crossing diabatic levels is assumed to be zero, as the coupling between levels 2 and 6 in Fig.~\ref{b1-fig}(b). If more than two levels intersect at one point, then nonzero couplings between different levels are additionally marked, as between levels 3 and 5 or 5 and 6 in Fig.~\ref{b1-fig}(b). 
Direct couplings between parallel levels are always considered zero, such as between levels 2 and 3 in Fig.~\ref{b1-fig}(a); this is included in the definition of the diabatic basis.  Using these rules, we can read off the Hamiltonian from such a diagram. For example, for Fig.~\ref{b1-fig}(a) the Hamiltonian is a 3$\times$3 matrix
\be
\hat{H}_{\rm DO} =\left( 
\begin{array}{ccc}
 t/2 & g_1&  g_2 \\
g_1 & e_1-t/2 & 0 \\
g_{2} &0 &e_2-t/2
\end{array} \right), \quad e_1<e_2,
\label{h-do3}
\ee 
and the Hamiltonian of the model in Fig.~\ref{b1-fig}(b) is a 6$\times$6 matrix:
 \begin{equation}
\hat{H}_{\rm 2B}=\left( 
\begin{array}{cccccc}
2e_2 -t & 0 &0 & \sqrt{2}g_2 & 0  & 0\\
0 & e_1+e_2  -t & 0 &   g_1         & g_2 &0 \\
0 & 0  & 2e_1 - t  & 0 & \sqrt{2}g_1 & 0\\
\sqrt{2}g_2  & g_1 &	0 & e_2    & 0 &	\sqrt{2}g_2 \\			
 0 &  g_2& \sqrt{2}g_1  &        0& e_1& \sqrt{2}g_1 \\
0 & 0 & 0&  \sqrt{2}g_2  &  \sqrt{2}g_1    & t
\end{array}
\right), \quad e_1<e_2.
\label{do-b2-ham}
\end{equation}

 Model (\ref{h-do3}) is known as the 3-state Demkov-Osherov (DO) model \cite{be,constraints,do}. It is one of the first  MLZ models that were solved by methods of complex analysis \cite{do}. Model (\ref{do-b2-ham}) is also solvable as it is a bosonic extension of the three-state DO model in the sense that it is realized by a quadratic Hamiltonian of three interacting bosonic modes:
\be
\hat{H}= \frac{1}{2}  t (\hat{a}^{\dg} \hat{a}-\hat{b}^{\dg} \hat{b} -\hat{c}^{\dg} \hat{c}) +e_1 \hat{b}^{\dg} \hat{b} + e_2 \hat{c}^{\dg} \hat{c} +\left(g_1\hat{a}^{\dg} \hat{b} +g_2\hat{a}^{\dg} \hat{c} +{\rm h. c.}\right),
\label{ham-b1}
\ee
where $\hat{a}$, $\hat{b}$ and $\hat{c}$ are bosonic annihilation operators.
This Hamiltonian conserves the number of bosons, so if we look at the two-boson sector of this model then in the basis
\begin{eqnarray}
\nonumber |1\ra &\equiv& \frac{ (\hat{c}^{\dg})^2}{\sqrt{2}} |0\ra, \quad |2\ra \equiv \hat{b}^{\dg} \hat{c}^{\dg} |0\ra, \quad |3\ra \equiv \frac{ (\hat{b}^{\dg})^2}{\sqrt{2}} |0\ra,  \\
|4\ra &\equiv& \hat{c}^{\dg} \hat{a}^{\dg} |0\ra, \quad |5\ra \equiv \hat{b}^{\dg} \hat{a}^{\dg} |0\ra, \quad |6\ra \equiv \frac{ (\hat{a}^{\dg})^2}{\sqrt{2}}|0\ra, 
\label{basis-6}
\end{eqnarray}
where $|0\ra$ is the vacuum state,
the Hamiltonian (\ref{ham-b1}) has the matrix form (\ref{do-b2-ham}). On the other hand, if we switch to the Heisenberg picture, then the quadratic Hamiltonian (\ref{ham-b1}) leads to  evolution  of operators
\be
i\frac{d}{dt}\left(
\begin{array}{c}
\hat{a}\\
\hat{b}\\
\hat{c}
\end{array}\right)
=\hat{H}_{\rm DO} \left(
\begin{array}{c}
\hat{a}\\
\hat{b}\\
\hat{c}
\end{array}\right),
\label{op-ev1}
\ee
 where $\hat{H}_{\rm DO}$ is the same as in (\ref{h-do3}). Let $S_{ij}$, $i,j\in 1,2,3$, be known scattering matrix elements of the DO model. Then we also have  relations among operators, such as $\hat{a}(+\infty)=S_{11} \hat{a}(-\infty)+S_{12} \hat{b}(-\infty)+S_{13} \hat{c}(-\infty)$ e.t.c..  One can use them
 to connect correlators of an arbitrary number of bosonic operators at $t=+\infty$ with correlators at $t=-\infty$, from which one can derive transition probabilities in the  model (\ref{do-b2-ham}). 
 This procedure was designed in \cite{multiparticle}. It was applied  to solve the model (\ref{do-b2-ham})  in \cite{jipdi}.

\subsection{Integrability Conditions}
\label{sec-IC}
The search for new solvable MLZ models by methods of complex analysis and multiparticle generalizations, as described above, has achieved only limited success. 
Recently, a much bigger class of solvable MLZ systems was discovered with a different approach based on ICs \cite{six-LZ,four-LZ,cQED-LZ}.  

The simplest way to introduce ICs is by using the diabatic level diagram. Let us say that a path in this diagram is {\it closed}  if it   goes along diabatic levels to produce a closed loop such that switching levels along this path is allowed only at level crossings with nonzero couplings. An example of such a path in Fig.~\ref{b1-fig}(b) is the one that starts at the crossing of levels 3, 5 and 6, then goes along level 6 forward in time, switches to level 4 at crossing with levels 4 and 1, then goes backward in time to level 2; then switches to level 5, and then goes backward in time to  return to the original crossing point with levels 3 and 6.  

We define the area inside such a closed path as the sum of areas of enclosed plaquettes in the diabatic level diagram counting clockwise and counterclockwise enclosed areas with opposite signs. We also introduce the notion of a {\it Hamiltonian projected to a subset of levels}. This Hamiltonian is obtained by setting couplings of given levels with all other states to zero and restricting the considered phase space only to the resulting evolution within the given subset of levels. 
ICs then say that for the full integrability of a MLZ model the following two conditions should be satisfied:
 
(i)  All closed paths of the graph in the diabatic level diagram should enclose zero areas. 

(ii) If two or several diabatic levels intersect at one point, and if there is no coupling path  that connects two diabatic states within the Hamiltonian projected to this subspace of  levels, then there is a corresponding pair of adiabatic energies of the full Hamiltonian that experiences exact crossing at all sufficiently small but finite values of coupling parameters.

For example, the DO model satisfies condition (i) trivially because the graph in Fig.~\ref{b1-fig}(a) is a simple tree so any closed path on  it has zero area naturally. In this model, there are also no diabatic level crossing points without direct coupling between levels, so condition (ii) is also satisfied trivially.

The case of the model in Fig.~\ref{b1-fig}(b)  is less trivial. It is easy to check that the triangle enclosed by levels 2, 4 and 6 is equal to the triangle enclosed by levels 2, 5 and 6, and that there is a single independent closed path in this level diagram that has to enclose one of these triangles clockwise and another one counterclockwise. So, the area of this path is zero, which proves condition (i). 

We will not prove validity of (ii) completely analytically because it is generally a complex task while the lack of minigaps at eigenvalue crossing points can be visualized and then tested numerically very quickly with an approach that was described in Ref.~\cite{cQED-LZ}. For example, to make the presence of exact level crossings visually obvious, we diagonalize the Hamiltonian (\ref{do-b2-ham}) and plot its eigenvalues as function of time in Fig.~\ref{b1-fig}(c). This figure shows three pairs of adiabatic level crossings that correspond to intersections of diabatic levels in Fig.~\ref{b1-fig}(b) with zero pairwise couplings, i.e. the pairs (3,4), (2,6) and (1,5), which proves property (ii) for this model. 

There are two more pairs of levels without direct couplings, namely (3,6) and (1,6) that, however, do not lead to exact crossings  of adiabatic levels. This is because these levels  cross simultaneously with other levels, respectively 5 and 4, to which they are coupled directly. So, if the Hamiltonian is projected to  three such simultaneously crossing levels the phase space does not split into disjoined subspaces. Condition (ii) allows such crossings to happen without emergence of exact adiabatic level crossings.

Finally, we note that our definition of ICs (i-ii) is slightly different from the one that was originally suggested in \cite{six-LZ}. Condition (i) here is more restrictive than just assumption of equality between dynamic phases of interfering semiclassical trajectories.  This would make no difference for all previously found solvable models. However, in appendix~\ref{a-counter}, we discuss an example of a model that breaks the condition (i) but does not break its counterpart in Ref.~\cite{six-LZ}. Our numerical tests showed that, at least by the standard semiclassical ansatz, this model is no longer  solvable. So the new version of condition (i) should be preferred. In appendix~\ref{sec:MLZ-graph} we also point that the new definition of the first integrability condition raises interesting analogies with properties of WKB quantized classically integrable multidimensional systems.
 On the other hand, our definition of condition (ii) is less restrictive than in Ref.~\cite{six-LZ} because now we allow level diagrams with more than two diabatic levels to cross at one point, as it is the case in Fig.~\ref{b1-fig}(b). 
\subsection{Solution in the form of the semiclassical ansatz } 
\label{sub-ansatz}
If ICs are satisfied then one can obtain scattering amplitudes of the model in the form of a  product of simple matrices. This form of the solution is called the matrix product ansatz or simply the {\it semiclassical ansatz}. The name follows from the fact that this ansatz coincides with prediction of the semiclassical approximation in which nonadiabatic processes near different crossing points happen independently of each other.  
 Assuming such  real values of all couplings, semiclassical ansatz is generated as follows \cite{six-LZ}: 

1)  We  list all level crossing points of the diabatic level diagram in chronological order of their appearance. 

2) For each crossing point, we identify the projected Hamiltonian that is obtained by keeping only diabatic levels that cross at this point and assuming that couplings to all other levels are zero. 

3) For each projected Hamiltonian, we determine the {\it projected scattering matrix}, i.e. the scattering matrix of MLZ model described by the projected Hamiltonian. We augment this scattering matrix then to the full phase space of the original model by assuming that other states simply do not experience interactions. 

4) The scattering matrix of the full model is then the chronologically ordered product of projected scattering matrices at all encountered crossing points.

 
5) Final transition probabilities are obtained by taking the absolute value squared of elements of this scattering matrix.

An equivalent formulation of the semiclassical ansatz can be done in terms of the {\it semiclassical trajectories} \cite{six-LZ} that are the paths that go forward in time along the diabatic levels, so that they start and end at, respectively, initially and finally populated levels. Each time such a path goes through a crossing point, one should prescribe a specific amplitude factor that this point contributes to the full trajectory amplitude as described in \cite{six-LZ}. The final amplitude of the transition is then the sum of amplitudes of all semiclassical trajectories connecting the given pair of states. 
 
Here we note that all parameters of  scattering matrices at individual crossing points are often not needed to be known. If there are no pairs of semiclassical trajectories that interfere with each other  in the full level diagram, then one can skip the step 4) and write the final transition probability matrix as the product of individual transition probability matrices at each crossing point.

 It was also found that, in the case when interference of trajectories matters, only trivial phase factors of elements of projected scattering matrices should be included in order to determine final transition probabilities \cite{six-LZ}. 
For example, if only two levels intersect at a given crossing point then the projected Hamiltonian at this crossing is the one of the two-state  Landau-Zener model \cite{book-LZ} with some coupling $g$ and crossing level slopes $\beta_{1}$ and $\beta_2$:
\begin{equation}
\hat{H}_{\rm LZ}=\left( 
\begin{array}{cc}
\beta_1 t&g \\
 g  & \beta_2 t\\
\end{array}
\right).
\label{lz-ham1}
\end{equation} 
 In all solvable models with a finite number of states, nonzero coupling constants can always be chosen real but not always positive.
The  simplified scattering matrix that corresponds to the projected two-state Hamiltonian  (\ref{lz-ham1}) is then known \cite{six-LZ}. It depends on the sign of $g$:
\begin{equation}
\hat{S}_{\rm LZ}=\left( 
\begin{array}{cc}
\sqrt{p} &\pm i\sqrt{1-p} \\
\pm i\sqrt{1-p}   & \sqrt{p}\\
\end{array}
\right), \quad  p=e^{-2\pi g^2/|\beta_1-\beta_2|},
\label{lz-s1}
\end{equation} 
 where sign ($\pm$) is the same as the sign of the coupling constant $g$. Note that the matrix (\ref{lz-s1}) does not include dynamic coupling-dependent phases of the full scattering matrix of the Landau-Zener model. The reason is that in all exactly solved models such phases factorize in the final result and do not influence final transition probabilities. Therefore, in what follows, we will use only such reduced versions of scattering matrices in our intermediate calculations, and verify validity of this choice at the end numerically.

\subsection{Examples of semiclassical ansatz solutions}
Let us consider the DO model (\ref{h-do3}) as an example. Since there is no interference between different paths connecting initial and final states, we can work directly with transition probability matrices that describe each crossing point. Let us define parameters
\be
p_{1,2}=e^{-2\pi g_{1,2}^2}, \quad q_{1,2}=1-p_{1,2}.
\label{pq1}
\ee 
There are two crossing points with nonzero couplings. In each case, the Hamiltonians projected to the crossing points are just the two state ones of the form (\ref{lz-ham1}), so we can introduce matrices of transition probabilities for evolution with such projected Hamiltonians:  
\begin{equation}
\hat{P}_1=\left( 
\begin{array}{ccc}
p_1 & q_1 & 0\\
q_1 & p_1& 0 \\
0 & 0  & 1\\
\end{array}
\right), \quad
\hat{P}_2=\left( 
\begin{array}{ccc}
p_2 & 0 & q_2\\
0 & 1& 0 \\
q_2 & 0 & p_2\\
\end{array}
\right).
\label{p-do3}
\end{equation} 
The full transition probability of the three-state DO model is then the product of these matrices taken in  chronological order:
\begin{equation}
\hat{P}_{\rm DO} = \hat{P}_2 \hat{P}_{1}.
\label{do-p2}
\ee

Let us now  turn to the model   (\ref{do-b2-ham}).
As in any solvable model, its scattering matrix factorizes in chronologically ordered contributions from all diabatic level crossing points with nonzero couplings. There is again no path interference in Fig.~\ref{b1-fig}(b) of the model  (\ref{do-b2-ham}), so the full transition probability matrix can be directly written as a product of four
 transition probability matrices describing four crossing points:
\be
\hat{P}_{\rm 2B} = \hat{P}_4 \hat{P}_3 \hat{P}_2 \hat{P}_1.
\label{fact1}
\ee
The matrix $\hat{P}_1$ here describes transitions at simultaneous crossing  of levels 3, 5, and 6. Projected to the subspace of these three levels Hamiltonian  is a 3$\times$3 matrix. After the  shift of the crossing point to $t=0$, this Hamiltonian reads
\begin{equation}
\hat{H}_1=\left( 
\begin{array}{ccc}
-t & \sqrt{2}g_1 & 0\\
\sqrt{2}g_1 & 0& \sqrt{2}g_1 \\
0 & \sqrt{2}g_1  & t\\
\end{array}
\right).
\label{do-b2-ham2}
\end{equation} 
It describes the model of a spin-1 in a linearly time-dependent magnetic field. The transition probability matrix for the model (\ref{do-b2-ham2}) is known since the time of Majorana \cite{maj}:
\begin{equation}
\hat{P}_1=\left( 
\begin{array}{ccc}
p_1^2& 2p_1 q_1& q_1^2 \\
2p_1q_1&(1-2p_1)^2 & 2p_1q_1 \\
q_1^2& 2p_1q_1& p_1^2 \\
\end{array}
\right),
\label{do-b2-ham3}
\end{equation} 
or in the basis of all six states of the original model
\begin{equation}
\hat{P}_1=\left( 
\begin{array}{cccccc}
1& 0 &0 &0& 0  & 0\\
0 & 1 & 0 &  0         & 0 &0 \\
0 & 0  & p_1^2 & 0 & 2p_1q_1 & q_1^2\\
0  & 0&	0 &   1 & 0 &	0 \\			
 0 &  0&2p_1q_1  & 0    & (1-2p_1)^2& 2p_1q_1 \\
0 & 0 &   q_1^2  & 0 &2p_1q_1   & p_1^2
\end{array}
\right).
\label{do-b2-ham4}
\end{equation} 
Similarly, $\hat{P}_2$ is a unit matrix except the 2$\times$2 block that corresponds to transitions between levels 2 and 4. Corresponding probabilities are given by the standard two-level Landau-Zener formula:
\be
\hat{P}_2=\left( 
\begin{array}{cccccc}
1& 0 &0 &0& 0  & 0\\
0 & p_1 & 0 &  q_1       & 0 &0 \\
0 & 0  & 1 & 0 & 0 & 0\\
0  & q_1 &	0&   p_1 & 0 &	0 \\			
 0 &  0&0 & 0    & 1& 0 \\
0 & 0 &   0 & 0 &0   & 1
\end{array}
\right).
\label{do-b2-ham5}
\end{equation} 
Matrices $\hat{P}_3$ and $\hat{P}_4$ are constructed analogously:
\be
\hat{P}_3=\left( 
\begin{array}{cccccc}
1& 0 &0 &0& 0  & 0\\
0 & p_2 & 0 &  0       & q_2 &0 \\
0 & 0  & 1 & 0 & 0 & 0\\
0  &0&	0&   1 & 0 &	0 \\			
 0 &  q_2&0 & 0    & p_2& 0 \\
0 & 0 &   0 & 0 &0   & 1
\end{array}
\right), \quad
\hat{P}_4=\left( 
\begin{array}{cccccc}
p_2^2& 0 &0 &2p_2q_2& 0  & q_2^2\\
0  & 1 &  0 &0       & 0 &0 \\
0 & 0  & 1 & 0 & 0 & 0\\
2p_2q_2  & 0 &	0&   (1-2p_2)^2& 0 &	2p_2q_2\\			
 0 &  0&0 & 0    & 1& 0 \\
q_2^2 & 0 &   0 & 2p_2q_2 &0   & p_2^2
\end{array}
\right).
\label{do-b2-ham5}
\end{equation} 
Such an approach to obtain the full transition probability matrix (\ref{fact1}) of the model (\ref{do-b2-ham}) is much more straightforward than dealing with correlators of bosonic operators and relating them to transition probabilities.

\section{General 4-state system satisfying integrability conditions}
\label{s-four}
\begin{figure}
 \scalebox{0.35}[0.35]{\includegraphics{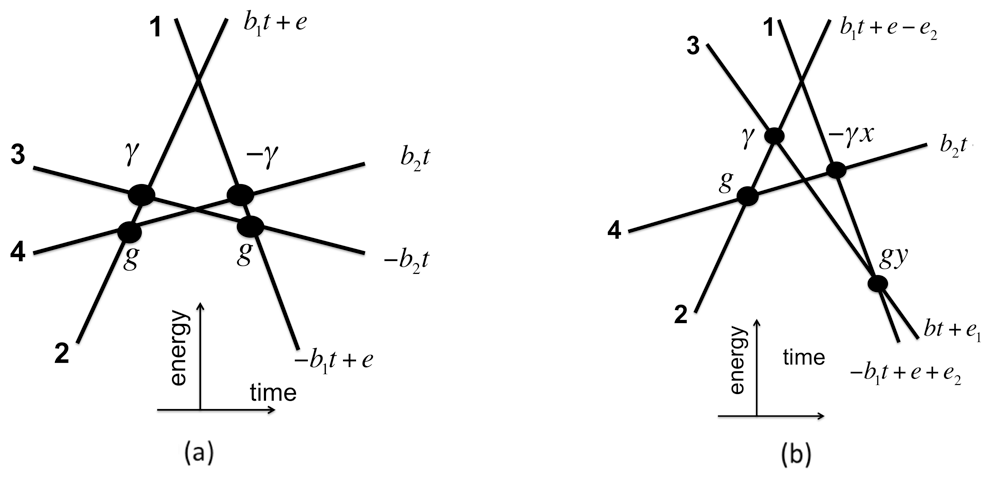}}
\hspace{-2mm}\vspace{-4mm}   
\caption{ Diabatic level diagram for Hamiltonians in (a) Eq.~(\ref{four-ham0}) that describes a known solvable model, and (b) Eq.~(\ref{four-ham1}) that is obtained by distortion of the model (a) such that the diabatic level diagram has the same topology but depends on additional parameters $b$, $e_1$, $e_2$, $x$ and $y$. }
\label{distort-fig1}
\end{figure}
\subsection{Distorting previously solved model}
 In previous section, we showed that if a model satisfies ICs then derivation of its transition probability matrix is straightforward. 
 However, the main challenge usually is to identify systems satisfying these conditions. 
 Here we propose an approach that is based on our new conjecture stating that { already solved MLZ 
 models are actually special limits of some more complex solvable models}.  
 
For example, we know  that solution for spin-1 in a 
 linearly time-dependent field \cite{maj} was later found to be a special case of the 
 three-state bow-tie model \cite{hioe}, which in turn was generalized later to the 
 solvable multistate version \cite{bow-tie}, and
 eventually to the solvable generalized bow-tie model \cite{gbow-tie}. So, it is reasonable to expect that 
 other known exact results in MLZ theory have similar generalizations. 
  
The utility of this conjecture is that an already solved system provides  the graph of diabatic levels with required {\it topology}. Here we say that  graphs of diabatic levels, or corresponding models, have the same topology if  the graphs have arbitrarily different parameters (couplings, level slopes) but the same set of level crossing points and types of level crossings. The latter means that two conditions are satisfied. First, levels with the same indexes in both graphs cross in the same points. Second, choices of parameters of corresponding Hamiltonians should be such that points leading to exact eigenvalue crossings should be the same in both graphs. One consequence of the latter rule is that,   for a pairwise level crossing, if the direct coupling between corresponding two levels is zero in one model it should  be also zero in the topologically equivalent model. 

We can use the graph of the previously solved MLZ model to design a more complex solvable model by
 {\it distorting} this graph, which means assuming a more general choice of level slopes and coupling parameters while preserving the model's 
  topology. 
  Generally, assuming arbitrary values of level slopes and couplings would violate ICs, however, for an available graph, we can reimpose such conditions 
in the form of relations within the new set of parameters. We are going to show that  the resulting model is generally richer  than the original one in the sense of the larger number of independent parameters. Moreover, it usually contains the original model as a special case. 

As the first example, 
let us consider a simple four-state model that was shown to satisfy  ICs and solved in \cite{four-LZ}.  It has the Hamiltonian
\begin{equation}
\hat{H}(t)=\left( 
\begin{array}{cccc}
-b_1t+e & 0 & g & -\gamma \\
0 & b_1 t  +e&\gamma &   g         \\
g & \gamma  & -b_2t & 0 \\
-\gamma   & g &	0 & b_2t \\			
\end{array}
\right),
\label{four-ham0}
\end{equation} 
and its level diagram is plotted in Fig.~\ref{distort-fig1}(a). Exact validity of the semiclassical ansatz for this model was rigrously proved in 
\cite{constraints}.

Let us now deform this graph as shown in Fig.~\ref{distort-fig1}(b) so that the new graph still has two crossings of diabatic levels without direct couplings.  All nonzero couplings can now be  considered different, which leads to the following Hamiltonian:
\begin{equation}
\hat{H}(t)=\left( 
\begin{array}{cccc}
-b_1t+e+e_2 & 0 & g y & -\gamma x \\
0 & b_1 t  +e-e_2&\gamma &   g         \\
gy & \gamma  & bt+e_1 & 0 \\
-\gamma x  & g &	0 & b_2t \\			
\end{array}
\right),
\label{four-ham1}
\end{equation} 
with some constants $b_1,b_2,b,e,e_1,e_2,x,y,g,\gamma$. Here we used   the gauge freedom \cite{be} so that slopes of levels 1 and 2 differ only by sign, and we set constant diagonal element of level 4 to  zero.

Parameter  $e_2$ can be  set to zero by time shift $t \rar t+e_2/b_1$ that
does not affect transition probabilities for evolution from $-\infty$ to $+\infty$. After this, we can make gauge transformation
of diabatic state amplitudes $a_i \rar e^{-itb_2 e_2/b_1} a_i$, $i=1,\ldots,4$, which results in $e_2=0$ at redefined parameters $e\rar e-b_2 e_2/b_1$ and $e_1 \rar e_1+(b-b_2)e_2/b_1$. 
We  also note that the gauge freedom to choose zero energy point and the arbitrariness of choosing indexes of states allow us to assume that 
\be
 b_1>b_2, \quad {\rm and} \quad b<b_2.
 \label{bconst3}
 \ee
  Otherwise, we can just rename indexes, followed by proper renaming of constant parameters and obtain the same Hamiltonian at desired relations (\ref{bconst3}).  
Therefore, in what follows, we will assume that $e_2=0$  with constraints in (\ref{bconst3}), unless specially stated.



Let $t_{ij}$ be the time moment of diabatic crossing between levels $i$ and $j$. For crossings without direct couplings we have:
\be
t_{12}=0, \quad t_{34}= e_1/(b_2-b),
\label{times1}
\ee
and for time moments of avoided crossings (i.e. moments of crossing of corresponding diabatic energy levels):
\be
t_{13}=\frac{e-e_1}{b_1+b}, \quad t_{14}= \frac{e}{b_1+b_2}, \quad t_{23}=-\frac{e-e_1}{b_1-b}, \quad t_{24}=-\frac{e}{b_1-b_2}.
\label{times2}
\ee

Let us now consider the condition on that the area inside the closed path in the graph in Fig.~\ref{distort-fig1}(b) is zero. This area is the difference between areas swept over the time axis by two trajectories that connect crossing points at time moments $t_{24}$ and $t_{13}$. One such a trajectory  starts at $t_{24}$, then  stays on level 2 till time $t_{23}$, and then goes along level 3 till time $t_{13}$. 
The area that this  trajectory sweeps over the time axis is 
$$
A_1=\int_{t_{24}}^{t_{23}}  (b_1t +e) \, dt +\int_{t_{23}}^{t_{13}}  (bt +e_1) \, dt. 
$$
The second trajectory starts at $t_{24}$, then  stays on level 4 till time $t_{14}$, and then goes along level 1 till time $t_{13}$. Explicitly, for two trajectories we have then 
\be
A_1=\frac{b_1}{2}(t_{23}^2-t_{24}^2)+\frac{b}{2} (t_{13}^2-t_{23}^2) +e(t_{23}-t_{24}) +e_1(t_{13}-t_{23}), 
\label{A1}
\ee
\be
A_2=\frac{b_2}{2}(t_{14}^2-t_{24}^2)-\frac{b_1}{2} (t_{13}^2-t_{14}^2) +e(t_{13}-t_{14}).
\label{A2}
\ee
Setting $A_1-A_2=0$ we find equation that  parameters of the model satisfy in order to realize IC (i). In terms of $e_1$ this equation is quadratic with two roots:
\be
e_1^{\pm}= e \pm |e| \sqrt{ \frac{ b_1^2-b^2}{b_1^2-b_2^2  }}.
\label{epm1}
\ee
Equation~(\ref{epm1}) shows that only values of $b$ in the range $-b_1 <b <b_1$ lead to physical  real valued $e_1^{\pm}$.

Our parameters do not satisfy IC (ii) yet. Generally, an analytical proof that some crossing of diabatic levels corresponds to an exact crossing point of adiabatic energy levels is a complex problem. However, there are perturbative conditions that must be satisfied by parameters if there is an exact crossing point. We will derive such constraints analytically and then check numerically whether imposing them is sufficient to satisfy condition (ii).

\begin{figure}
\scalebox{0.32}[0.32]{\includegraphics{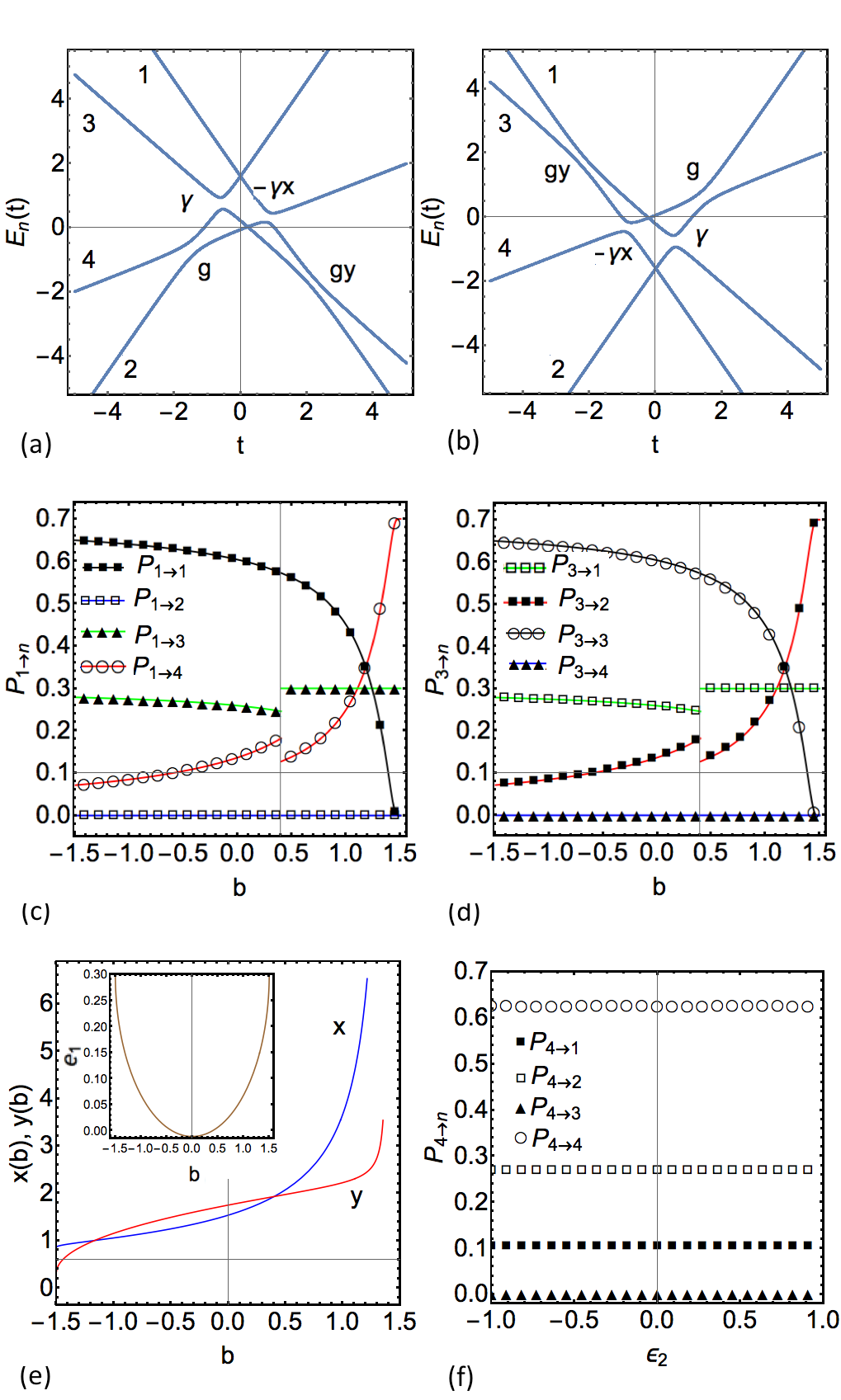}}
\hspace{-2mm}\vspace{-4mm}   
\caption{Characteristics of the phase with path interference. (a-b) Adiabatic energy levels  at, respectively, $e>0$ and $e<0$ (with, respectively, $e_1=e_{1}^{-}$, and $e_1=e_{1}^{+}$). Parameters in both cases: 
$ b_1=1.5,\, b_2=0.4,\, g=0.25,\, \gamma=0.75g,\, e_2=0$, and (a) $ b=-0.35,\, e=1.55$, (b) $b=-0.1,\, e=-1.55$. Avoided crossings are marked by corresponding couplings between diabatic levels. (c-d) Test
of the validity of Eqs.~(\ref{tprob1})-(\ref{tprob2}) for different elements of the transition probability matrix as functions of parameter $b$. Solid curves are theoretical predictions and discrete points are results of numerical simulation of evolution from $t=-2000$ to $t=2000$ with a discrete time step $dt=0.0005$. Numerical algorithm is described in Ref.~\cite{cQED-LZ}. Choice of parameters:  $e=0.3,\, e_2=0.1,\,b_1=1.5,\, b_2=0.4,\, g=0.25,\, \gamma=0.75g$.
(e) Dependence of coupling asymmetry parameters  $x$ and $y$, as well as $e_1$ in the inset, on $b$ for the same other parameters as in (c). (f) Test of independence of transition probabilities of $e_2$. 
Other parameters are as in (c) with $b=1.25$. Parameters $e_1$, $x$, $y$ are always tuned to satisfy integrability conditions, including at nonzero $e_2$.
}
\label{b22-fig}
\end{figure}

Let $E_k$, $k=1,\ldots, 4$ be diabatic energies of levels, respectively, $1,\ldots,4$ at time $t_{12}$. We also define energy gaps 
\be
E_{13}\equiv E_1-E_3=e-e_1, \quad E_{14}\equiv E_1-E_4 = e.
\label{eij1}
\ee
Similarly, let $E_k'$ be energies of levels at time $t_{34}$. We have then 
\be
E_{32}'\equiv E_{3}'-E_{2}' =-e-\frac{e_1(b_1-b_2)}{b_2-b} , \quad E_{31}'\equiv E_{3}'-E_{1}' =-e+\frac{e_1(b_1+b_2)}{b_2-b}.
\label{eij2}
\ee
Let us now treat all couplings as small. 
The necessary condition for the crossing point  between levels 1 and 2 to be exact is that coupling between these diabatic states remains zero at 2nd order perturbation in $\gamma$ and $g$. This leads to the constraint 
$$
-\frac{\gamma x g}{E_{14}}+ \frac{\gamma g y}{E_{13}}=0.
$$
Similarly, coupling between levels 3 and 4 should be zero at 2nd order perturbation. This gives the constraint 
$$
xy=E_{31}'/E_{32}',
$$
which leads to 
\be
x=\pm \sqrt{\frac{E_{31}' E_{14} }{E_{32}'E_{13} }} , \quad y=x E_{13}/E_{14}.
\label{f-const3}
\ee
Thus, in order to satisfy ICs, there are three constraints in (\ref{epm1}) and (\ref{f-const3}) on parameters so that only  parameter $b$ remains independent among the newly introduced ones. 
The original model in Fig.~\ref{distort-fig1}(a) is recovered at $b=-b_2$ but other values of $b$ produce models that have not encountered previously in MLZ theory. 

\begin{figure}
\scalebox{0.32}[0.32]{\includegraphics{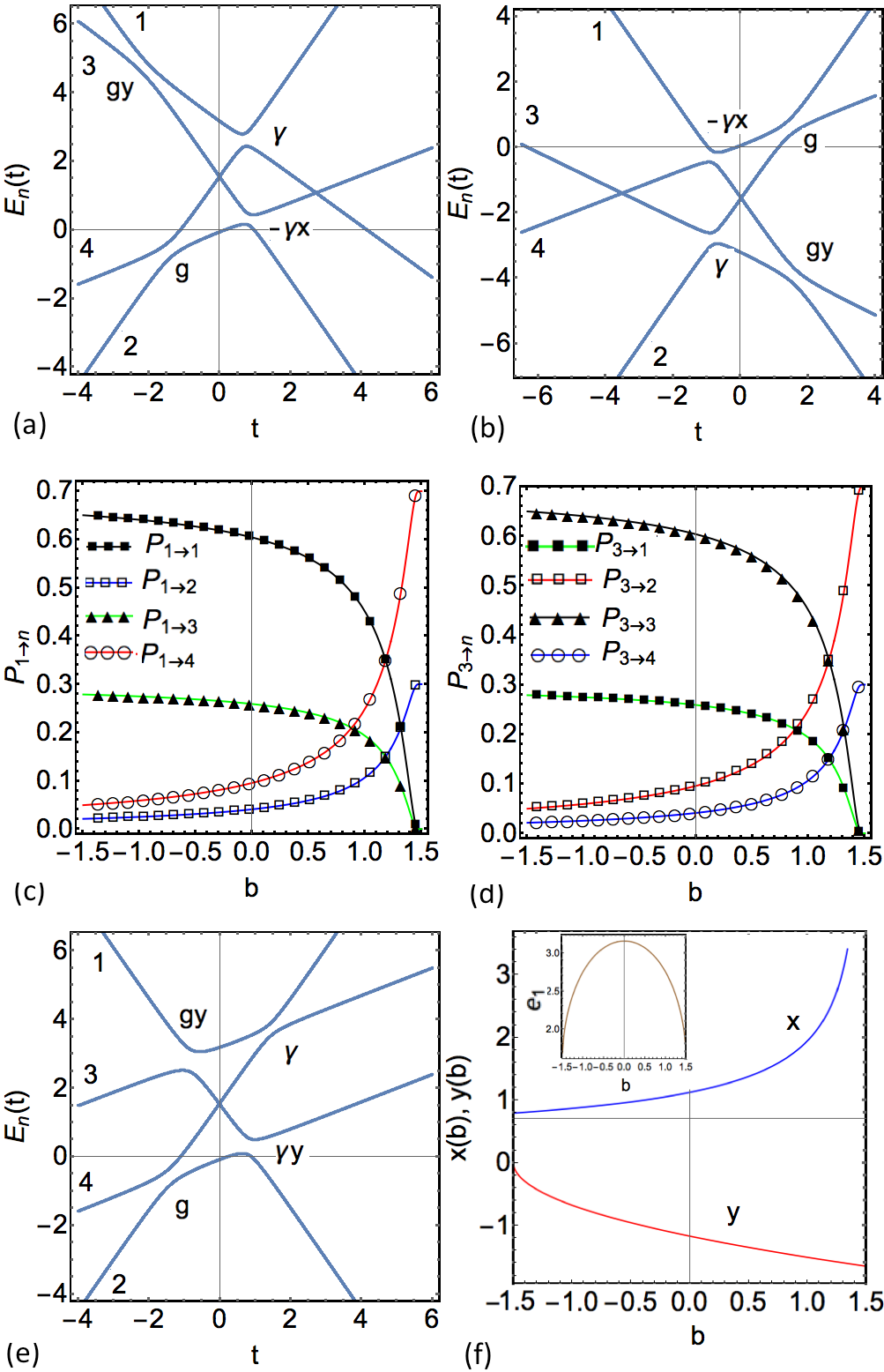}}
\hspace{-2mm}\vspace{-4mm}   
\caption{Characteristics of the phase without path interference. (a-b) Adiabatic energy levels  at, respectively, $e>0$ and $e<0$ (with, respectively, $e_1=e_{1}^{+}$, and 
$e_1=e_{1}^{-}$). Parameters in both cases are:
$ b_1=1.5,\, b_2=0.4,\, g=0.25,\, \gamma=0.75g,\, e_2=0$, and (a) $ b=-0.35,\, e=1.55$, (b) $b=-0.1,\, e=-1.55$. 
Avoided crossings are marked by corresponding coupling between diabatic levels. (c-d) Test
of the validity of Eq.~(\ref{tprob3}) for different elements of the transition probability matrix as functions of parameter $b$. 
Solid curves are theoretical predictions and discrete points are results of numerical simulation.
 Choice of constant parameters in (c) and (d) are as in, respectively, (a) and (b). (f) Eigenvalues of the Hamiltonian at $b=b_2$ and other parameters are as in (a). 
(f) Dependence of coupling asymmetry parameters  $x$ and $y$, as well as $e_1$ in the inset, on $b$ for the same other parameters as in (a).
}
\label{b2-fig2}
\end{figure}

Conditions (\ref{f-const3}) are necessary but not sufficient in order to guarantee the presence of exact crossing points near times $t_{12}$ and $t_{34}$. Therefore, we resort to numerical simulations for additional proof. Figures~\ref{b22-fig}(a-b) and Figs.~\ref{b2-fig2}(a-b) show results of explicit numerical calculation of the adiabatic energy levels (i.e., eigevnalues of the Hamiltonian as functions of time $t$) at different values of parameters. It is visually clear that there are, indeed, pairs of exact eigenvalue crossings in each figure when conditions (\ref{epm1}) and (\ref{f-const3}) are satisfied.

 Depending on the sign in Eq.~(\ref{epm1}) and parameter $e$, one can generate different graph types. Two types correspond to models in which  different semiclassical trajectories can interfere, while other two graph types  do not feature path interference.

\subsection{Phase with path interference}

 In Figs.~\ref{b22-fig}(a-b), we show adiabatic energies of the Hamiltonain in  cases when either (a) $e_1=e_1^{-}$ with $e>0$ or (b) $e_1=e_1^{+}$ with $e<0$. Geometries of these graphs are different by the relative positions of 
the exact crossing points. The common feature of both graphs is the possibility of more than one semiclassical trajectories that connect one diabatic state at $t=-\infty$ with another diabatic state at $t=+\infty$. 
In both cases we find
\be
x=\pm \sqrt{\frac{b_1+b_2}{b_1-b}}, \quad y=\pm \sqrt{\frac{b_1+b}{b_1-b_2}},
\label{f-const4}
\ee
with $x$ and $y$ having the same sign. 
Let us introduce 
\be
p_1=e^{-2\pi g^2/(b_1-b_2)}, \quad  p_2=e^{-2\pi \gamma^2/(b_1-b)},  \quad q_{1,2}=1-p_{1,2}.
\label{pq1}
\ee
Semiclassical ansatz calculations with parameters (\ref{f-const4}) are analogous to calculations in Ref.~\cite{four-LZ} that were done for the original model from Fig.~\ref{distort-fig1}(a). They lead  to the following simple form of the transition probability matrix:
\be
\hat{P}=\left(
\begin{array}{cccc}
p_1p_2 &0 &p_2q_1 &q_2\\
0 &p_1p_2  & q_2 &p_2q_1\\
p_2q_1& q_2 &p_1p_2 & 0 \\
q_2&  p_2 q_1& 0 & p_1p_2
\end{array} \right), \quad b_1>b_2>b,
\label{tprob1}
\ee
 \be
\hat{P}=\left(
\begin{array}{cccc}
p_1p_2 &0 &q_1&q_2p_1\\
0 &p_1p_2  & q_2p_1 &q_1\\
q_1& q_2p_1 &p_1p_2 & 0 \\
q_2p_1&  q_1& 0 & p_1p_2
\end{array} \right), \quad b_1>b>b_2.
\label{tprob2}
\ee
 
Figures~\ref{b22-fig}(c-d) compare solution  (\ref{tprob1})-(\ref{tprob2}) with results of direct numerical simulations at different values of parameter $b$. The agreement between analytical predictions and numerics is perfect for all values of parameters.  This leaves no doubt that semiclasscial ansatz prediction here is not an approximation but rather the exact result although, certainly, such an agreement with numerics cannot be accepted as a mathematically rigorous proof.  Behavior of parameters $x$ and $y$ is provided in Fig.~\ref{b22-fig}(e) in order 
to show that, in our numerical tests, these parameters  varied substantially. 
For generality, we assumed in  simulations that $e_2 \ne 0$, so we also show Fig.~\ref{b22-fig}(f) confirming that varying $e_2$ does not influence transition probabilities, as long as we adjust $e_1$ properly to satisfy integrability conditions.
 
Transition probabilities in Figs.~\ref{b22-fig}(c-d) show discontinuous behavior at $b=b_2$. This is not surprising because at $b=b_1$ we have $e_1=0$, so diabatic levels 3 and 2 become degenerate not at a single time moment but  during all the time. Transition probabilities in MLZ models are known to behave generally discontinuously  when parameters pass through values that create such permanent degeneracies \cite{do}.  
 
 Interestingly, solution (\ref{tprob1})-(\ref{tprob2}) for the model in Fig.~\ref{distort-fig1}(b) has exactly the same form as the solution for the original model in Fig.~\ref{distort-fig1}(a) with only minor generalization that allows free choice of parameter $b$. This hints to possible symmetry that relates arbitrary values of $b$ to the case with $b=-b_2$, which was rigorously proved in \cite{constraints}.

 \subsection{Phase without path interference}

 Let us now turn to phases such that (a) $e_1=e_1^{-}$ with $e<0$ or (b) $e_1=e_1^{+}$ with $e>0$ (see Eq.~(\ref{epm1})).
 Repeating perturbative calculations, we  find that in this case parameters $x$ and $y$ have opposite signs: 
 \be
x=\pm \sqrt{\frac{b_1+b_2}{b_1-b}}, \quad y=\mp \sqrt{\frac{b_1+b}{b_1-b_2}},
\label{f-const5}
\ee
i.e, negative couplings appear in pairs.  

In Figs.~\ref{b2-fig2}(a-b) we show possible patterns of adiabatic energy levels and demonstrate two exact crossing points. Apparently, there is no path interference in this case. In both cases
 we 
find a  solution 
  \be
\hat{P}=\left(
\begin{array}{cccc}
p_1p_2 &q_1q_2 &p_2q_1&p_1q_2\\
q_1q_2&p_1p_2  & p_1q_2 &p_2q_1\\
p_2q_1& p_1q_2 &p_1p_2 & q_1q_2 \\
p_1q_2&  p_2q_1& q_1q_2 & p_1p_2
\end{array} \right).
\label{tprob3}
\ee

Results of numerical tests of this prediction are shown in Figs.~\ref{b2-fig2}(c-d). Again, we find perfect agreement  between the semiclassical ansatz  and numerical predictions. In Figs.~\ref{b2-fig2}(e,f) we  show, respectively,
adiabatic levels at $b=b_2$ and dependence of parameters $x$, $y$, $e_1$ on $b$ that we used to satisfy ICs. 

 Several properties of solution (\ref{tprob3}) are to be mentioned. First, up to exchange of positions of some raws and columns, this transition probability matrix is a direct product of 2$\times$2 matrices 
$$
\left(
\begin{array}{cc}
p_1 &q_1 \\
q_1&p_1  
\end{array} \right) \quad {\rm and} \quad 
\left(
\begin{array}{cc}
p_2 &q_2 \\
q_2&p_2 
\end{array} \right),
$$
which are transition probability matrices of two two-state Landau-Zener models. 
In fact, the special case with $b=-b_2$ corresponds to the solvable model from Ref.~\cite{multiparticle}. This model is, indeed, constructed as a  direct product of two two-state  Hamiltonians. This fact also suggests that the case with arbitrary $b$ may be related  to the direct product of two decoupled two-state Landau-Zener systems.

Second interesting property is the absence of discontinuity at $b=b_2$. This follows from the fact that in this case we have $e_1=2e \ne 0$, i.e., there is no permanent degeneracy of two levels. 
At this point, we have $x=-y=\sqrt{(b_1+b_2)/(b_1-b_2)}$, and two levels are parallel to each other, as illustrated in Figs.~\ref{b2-fig2}(e). This case, however, is different from another known solvable model, called the four-state bow-tie model \cite{bow-tie}, that has the same geometry of level crossings but different constraints on coupling constants. We find it surprising that the four-state bow-tie model is not a special case of our model. This observation means that, when some of the diabatic levels are parallel to each other, one has more options to satisfy ICs.  Determining constraints on parameters in such cases is, however, more difficult because nontrivial constraints emerge then in higher order perturbation theory. We  consider such a model in the following section.





\section{Distorting Bosonic model}
\label{s-six}
Let us now consider the bosonic model represented by Fig.~\ref{b1-fig}(b) and look for possibility to distort it in order to generate a new solvable model. 

\subsection{Model with new couplings only}

\begin{figure}
\scalebox{0.35}[0.35]{\includegraphics{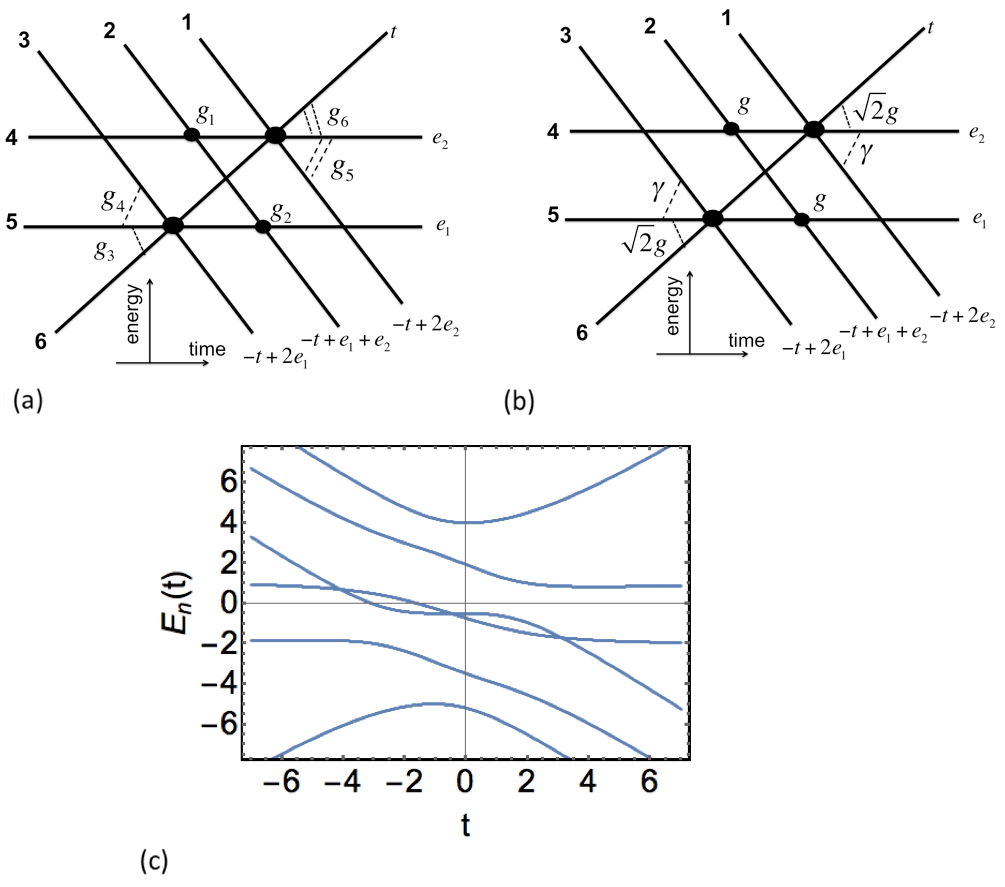}}
\hspace{-2mm}\vspace{-4mm}   
\caption{ Diabatic level diagram for (a) the Hamiltonian  that has the same structure as in Eq.~(\ref{do-b2-ham}) but with arbitrary choice of coupling constants (b) with specific choice, which is  different from the one in Eq.~(\ref{do-b2-ham}) but satisfying integrability conditions. (c) Three exact crossing points in spectrum of the Hamiltonian that corresponds to Fig.~\ref{b2-fig}(b) at parameter values $\gamma=1.123$, $g=1.5$, $e_1=1$, $e_2=-2$. }
\label{b2-fig}
\end{figure}

For simplicity, first, we will not change the slopes of diabatic levels at all and try only to find a different choice of coupling parameters that would satisfy condition (ii). 
According to the strategy that we described in previous section,  we augment the set of parameters to generate the model shown in 
Fig.~\ref{b2-fig}(a) with six unknown coupling parameters $g_1$-$g_6$. It is easy to check numerically that a random choice of these couplings leads to permanent gaps between all adiabatic energy levels, so this model is generally not integrable. 

To narrow the search for proper couplings, we treat coupling constants as small. Diabatic levels 2 and 6 cross at $t_{cr}= (e_1+e_2)/2$. 
Let $E_k$, $k=1,\ldots,6$ be diabatic energies of states at this time moment. We will need
$$
E_2=E_6=(e_2+e_1)/2, \quad E_4=e_2, \quad E_5=e_1.
$$
At second order in strength of parameters $g_i$, coupling between  levels 2 and 6 is then given by 
\be
v_{26}= \frac{g_1g_6}{E_2-E_4}+\frac{g_2g_3}{E_2-E_5}. 
\label{coupl1}
\ee
In order not to open the minigap between levels 2 and 6, this coupling should be zero. Since $E_2-E_4=-(E_2-E_5)$, this gives a constraint on couplings: 
\be
g_1g_6=g_2g_3.
\label{coupl2}
\ee
 Let us test now the ``no-coupling" condition for levels 3 and 4. These diabatic levels cross at $t_{cr} =2e_1-e_2$, and at this moment
 $$
 E_6=2e_1-e_2, \quad E_3=E_4=e_2, \quad E_5=e_1, \quad E_2=2e_2-e_1.
 $$ 
Generally, a minigap between  levels 3 and 4 would open as the third order correction:
\be
v_{34}= \frac{g_6g_3 g_4}{(E_4-E_6)(E_4-E_5)}+\frac{g_1g_2g_4}{(E_4-E_2)(E_4-E_5)}. 
\label{coupl3}
\ee 
By equating $v_{34}$ to zero we find  
 \be
g_3g_6=2g_1g_2.
\label{coupl4}
\ee
Taking ratio of Eqs.~(\ref{coupl2}) and (\ref{coupl4}) we find a constraint $g_3=\pm \sqrt{2} g_1$. Similarly, by considering crossing of levels 1 and 5, we find constraint $g_6=\pm \sqrt{2} g_2$. 

Such constraints are insufficient to determine values of parameters $g_4$ and $g_5$. This is expected because when a model has parallel diabatic levels, such levels do not give rise to adiabatic crossing between them, which reduces the number of independent constraints on coupling parameters.  So, for such cases, one should  develop higher order perturbative tests for existence of  exact level
crossings. Here, we will not go for  this because the already found constraints are sufficiently helpful. We just recall that presence of exact crossing points usually requires some symmetric choices of couplings. After trying several candidate sets, we found that the choice shown
in Fig.~\ref{b2-fig}(b) does lead to the desired exact eigenvalue crossings, as we prove in Fig.~\ref{b2-fig}(c).
 The Hamiltonian of this model reads:
\be
\hat{H} = \left( 
\begin{array}{cccccc}
-t+2e_2 & 0 &0 & \gamma &0&0\\
0 & -t+e_1+e_2 & 0 & g & g &0 \\
0&0&-t+2e_1 & 0& \gamma &0\\
\gamma & g& 0&e_2& 0 & \sqrt{2}g \\
0& g& \gamma & 0& e_1 & \sqrt{2} g \\
0&0&0& \sqrt{2}g & \sqrt{2}g & t
\end{array}
\right),
\label{bos-distH1}
\ee
with arbitrary real parameters $e_1$, $e_2$, $g$, and $\gamma$.

According to Sec.~\ref{sub-ansatz}, as in the original bosonic model, the matrix of transition probabilities factorizes as a product of four transition probability matrices describing  all crossing points with nonzero couplings.  
Consider first the crossing point of diabatic levels 3, 5, and 6. After the time  shift that places this crossing point at $t=0$, 
corresponding projected to the phase space of these levels Hamiltonian becomes a 3$\times$3 matrix:
\be
\hat{H}_1 = \left( 
\begin{array}{ccc}
-t & \gamma &0 \\
\gamma & 0& \sqrt{2}g  \\
0&\sqrt{2}g&t 
\end{array}
\right).
\label{bos-distH2}
\ee
Transition probabilities for evolution with this Hamiltonian are known because it is a special case of the Hamiltonian of the solvable bow-tie model. Let us define parameters
\be
p_{1}=e^{-2\pi g^2}, \quad p_2=e^{-\pi \gamma^2}, \quad q_{1,2}=1-p_{1,2}.
\label{pq1}
\ee
The transition probability matrix restricted to the crossing levels is then given by  \cite{hioe}
\begin{equation}
\hat{P}_1=\left( 
\begin{array}{ccc}
p_2^2& q_2(p_1+p_2) & q_1q_2 \\
q_2(p_1+p_2)&(1-p_1-p_2)^2 & q_1(p_1+p_2) \\
q_1q_2& q_1(p_1+p_2)& p_1^2 \\
\end{array}
\right),
\label{do-b2-ham3}
\end{equation} 
or in the basis of all six states
\begin{equation}
\hat{P}_1=\left( 
\begin{array}{cccccc}
1& 0 &0 &0& 0  & 0\\
0 & 1 & 0 &  0         & 0 &0 \\
0 & 0  & p_2^2 & 0 & q_2(p_1+p_2) & q_1q_2\\
0  & 0&	0 &   1 & 0 &	0 \\			
 0 &  0&q_2(p_1+p_2)  & 0    & (1-p_1-p_2)^2 & q_1(p_1+p_2) \\
0 & 0 &   q_1q_2  & 0 &q_1(p_1+p_2)   & p_1^2
\end{array}
\right).
\label{do-b2-ham41}
\end{equation} 
Similarly, $\hat{P}_2$ and $\hat{P}_3$ are unit matrices except the 2$\times$2 blocks that describe transitions near crossings of level pairs (2,4) and (2,5):
\be
\hat{P}_2=\left( 
\begin{array}{cccccc}
1& 0 &0 &0& 0  & 0\\
0 & p_1 & 0 &  q_1       & 0 &0 \\
0 & 0  & 1& 0 & 0 & 0\\
0  & q_1 &	0&   p_1 & 0 &	0 \\			
 0 &  0&0 & 0    & 1& 0 \\
0 & 0 &   0 & 0 &0   & 1
\end{array}
\right), \quad 
\hat{P}_3=\left( 
\begin{array}{cccccc}
1& 0 &0 &0& 0  & 0\\
0 & p_1 & 0 & 0       & q_1 &0 \\
0 & 0  & 1 & 0 & 0 & 0\\
0  & 0&	0&   1 & 0 &	0 \\			
 0 &  q_1 &0 & 0    & p_1& 0 \\
0 & 0 &   0 & 0 &0   & 1
\end{array}
\right),
\label{do-b2-ham51}
\end{equation} 
and transition probabilities at the crossing of  levels 1, 4, and 6 are again obtained from the solution of the three-state bow-tie model:
\begin{equation}
\hat{P}_4=\left( 
\begin{array}{cccccc}
p_2^2& 0 &0&q_2(p_1+p_2)& 0  & q_1q_2\\
0 & 1 & 0 &  0         & 0 &0 \\
0 & 0  & 1 & 0 & 0 & 0\\
q_2(p_1+p_2)  & 0&	0 &   (1-p_1-p_2)^2 & 0 &	q_1(p_1+p_2) \\			
 0 &  0&0   & 0    & 1& 0 \\
q_1q_2 & 0 &    0 & q_1(p_1+p_2) &0   & p_1^2
\end{array}
\right).
\label{do-b2-ham61}
\end{equation} 
According to Sec.~\ref{sub-ansatz}, the total transition probability matrix is given by $\hat{P}_{\rm fin} = \hat{P}_4\hat{P}_3\hat{P}_2\hat{P}_1$, or explicitly 
\be
\hat{P}_{\rm fin} =  \left( 
\begin{array}{cccccc}
p_2^2 & q_1q_2(p_1+p_2) & q_1^2q_2^2 & p_1q_2(p_1+p_2) & q_1^2q_2(p_1+p_2) & p_1^2q_1q_2\\
0& p_1^2 & q_1q_2(p_1+p_2) & p_1q_1& q_1(1-p_1-p_2)^2 & q_1^2(p_1+p_2)\\
0&0&p_2^2 & 0 &q_2(p_1+p_2) & q_1q_2 \\
q_2(p_1+p_2) & q_1(1-p_1-p_2)^2 & q_1^2q_2(p_1+p_2)& p_1(1-p_1-p_2)^2& q_1^2(p_1+p_2)^2 & p_1^2q_1 (p_1+p_2)\\
0& p_1q_1& p_1q_2(p_1+p_2) & q_1^2 & p_1 (1-p_1-p_2)^2 & p_1q_1 (p_1+p_2)\\
q_1q_2 & q_1^2(p_1+p_2)& q_1p_1^2q_2 & q_1 p_1 (p_1+p_2) & q_1p_1^2 (p_1+p_2) & p_1^4
\end{array}
\right).
\label{ptot1}
\ee
\begin{figure}
 \scalebox{0.30}[0.30]{\includegraphics{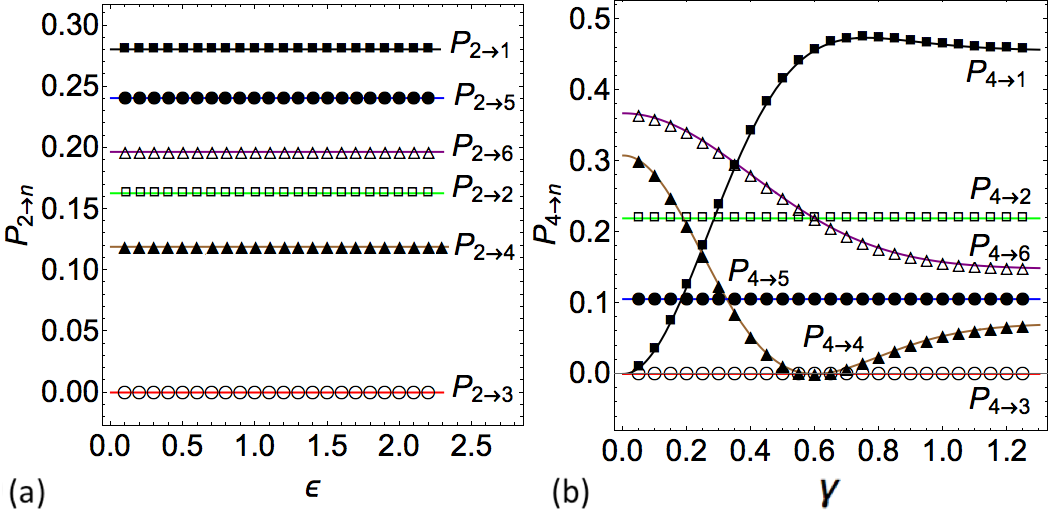}}
\hspace{-2mm}\vspace{-4mm}   
\caption{ Test of  validity of the transition probability matrix (\ref{ptot1}) for the model in Fig.~\ref{b2-fig}(b). Solid curves are analytical predictions and discrete points are results of numerical simulations for evolution from $t=-2000$ to $t=2000$ with a time step $dt=0.00005$. Algorithm is explained in Ref.~\cite{cQED-LZ}. (a)
Test of independence of transition probabilities  of  rescaled distances between parallel levels. Here $e_1= -\epsilon$, $e_2=1.3 \epsilon$. Evolution starts at level 2. Other parameters: $\gamma=0.779$, $g=0.38$. (b) Dependence of transition probabilities, starting from level 4, on coupling parameter $\gamma$. Other parameters: $e_1= -0.5$, $e_2=0.2$, $g=0.25$. }
\label{lz-six-fig2}
\end{figure}
In Fig.~(\ref{lz-six-fig2}) we test predictions of Eq.~(\ref{ptot1}) and find that agreement with numerical simulations is again excellent. Note that one of the signatures of integrability, which can be tested numerically, is that all transition probabilities are independent of rescaling the distances between parallel diabatic levels, as we show in Fig.~\ref{lz-six-fig2}(a). This property of the semiclassical ansatz is found in all solved models with parallel diabatic levels \cite{cQED-LZ,do,gbow-tie,no-go}. 

\subsection{Deformation leading to  path interference}
\begin{figure}
 \scalebox{0.30}[0.30]{\includegraphics{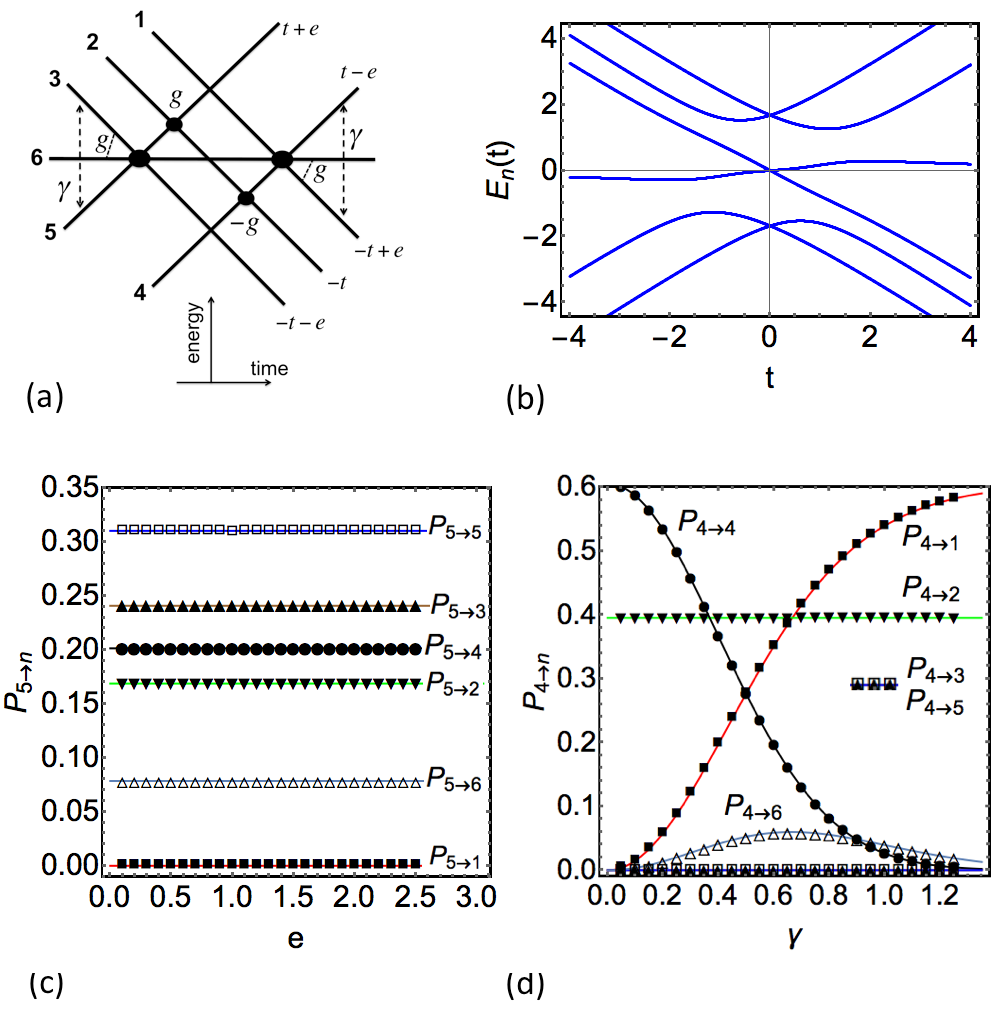}}
\hspace{-2mm}\vspace{-4mm}   
\caption{ (a) Diabatic level diagram of the model with the Hamiltonian (\ref{bos-distH4}). (b) 
Three exact crossing points of  Hamiltonian  (\ref{bos-distH4}) eigenvalues at $t=0$. Parameters: $e=1.$, $\gamma=1.$, $g=0.65$. (c-d) Tests of the validity of the semiclassical ansatz prediction (\ref{ptotS1}). Parameters: (c) $\gamma=0.35$, $g=0.5$; (d) $e=0.5$, $g=0.4$. Solid curves are analytical predictions and discrete points are numerical  results for evolution from $t=-2000$ to $t=2000$ with time step $dt=0.0005$.}
\label{add}
\end{figure}

Further extensions of the found solution can be searched among models with different level slopes but the same positions of nonzero couplings and keeping parallel levels to be parallel after the deformation. 
As physical meaning of such models is obscure, we will restrict ourselves to a particularly symmetric case in which we exchange the slope of level 6 with that of levels 4 and 5, as shown in Fig.~\ref{add}(a). We adjust distances between parallel diabatic levels
to satisfy IC (i) without changing the number and types of the crossing points. Couplings are found using IC (ii) by using the third order perturbation theory, just as in the previous subsection. 
The resulting model has the Hamiltonian
\be
\hat{H}' = \left( 
\begin{array}{cccccc}
-t+e & 0 &0 & \gamma &0&g\\
0 & -t & 0 & -g & g &0 \\
0&0&-t-e & 0& \gamma &g\\
\gamma & -g& 0&t-e& 0 & 0\\
0& g& \gamma & 0& t+e & 0 \\
g&0&g& 0 &0 & 0
\end{array}
\right),
\label{bos-distH4}
\ee 
which is parametrized by three independent parameters $e$, $g$, and $\gamma$.  
Figure~\ref{add}(b) shows numerical proof of the existence of three exact level crossings, as required by IC (ii) in this case. The fact that some of the couplings have different signs and that all three exact crossings happen simultaneously at $t=0$ indicates that this model can describe an odd-spin system \cite{four-LZ}. 

Examining Fig.~\ref{add}(a) we find that there is a possibility for interference between semiclassical trajectories, so we have to construct semiclassical ansatz from scattering rather than probability matrices of each crossing point. 
Consider first the crossing point of levels 3, 5, and 6. The effective Hamiltonian near this point is a 3$\times$3 matrix
\be
\hat{H}_1 = \left( 
\begin{array}{ccc}
-t & \gamma & g \\
\gamma & t & 0 \\
g& 0 & 0 
\end{array}
\right).
\label{bos-distH41}
\ee 
This is again the Hamiltonian of the three-state bow-tie model \cite{hioe} but with different from Eq.~(\ref{bos-distH2}) order of level slopes. 
Let 
\be
X=e^{-\pi \gamma^2/2}, \quad Y=e^{-\pi g^2}.
\label{XY}
\ee
The scattering matrix for the Hamiltonian (\ref{bos-distH41}) can be obtained by taking square roots of known transition probabilities (given explicitly in Ref.~\cite{hioe}), and adding factors $i$ taken to the power of the number of direct transitions that connect the levels:
\be
\hat{S}_1 = \left( 
\begin{array}{ccc}
X Y& i\sqrt{(1-X)(1+XY)} & i\sqrt{X(1-Y)(1+XY)}\\
i\sqrt{(1-X)(1+XY)}& X & -\sqrt{X(1-X)(1-Y)} \\
i\sqrt{X(1-Y)(1+XY)}& -\sqrt{X(1-X)(1-Y)} & 1-X+X Y 
\end{array}
\right).
\label{bos-distS41}
\ee 
Note that this scattering matrix is truncated so that we drop its dependence on coupling-dependent phase factors which do not change final transition probabilities in all known exactly solvable models.
Lifting $\hat{S}_1$ to the phase space of all six states, we get 
\begin{equation}
\hat{S}_1=\left( 
\begin{array}{cccccc}
1& 0 &0 &0& 0  & 0\\
0 & 1 & 0 &  0         & 0 &0 \\
0 & 0  & X Y & 0 & i\sqrt{(1-X)(1+XY)} & i\sqrt{X(1-Y)(1+XY)}\\
0  & 0&	0 &   1 & 0 &	0 \\			
 0 &  0&i\sqrt{(1-X)(1+XY)} & 0    &  X &  -\sqrt{X(1-X)(1-Y)}\\
0 & 0 &  i\sqrt{X(1-Y)(1+XY) } & 0 & -\sqrt{X^2(1-X)(1-Y)}  & 1-X+X Y 
\end{array}
\right).
\label{do-b2-S41}
\end{equation} 
Chronologically, the next two crossing points  correspond to pairwise transitions  between levels (2,5) and then (2,4).  Corresponding scattering matrices have the form (\ref{lz-s1}) with one subtlety: due to different signs of couplings near those intersections, off-diagonal elements of corresponding scattering matrices have opposite signs:
\be
\hat{S}_2=\left( 
\begin{array}{cccccc}
1& 0 &0 &0& 0  & 0\\
0 & Y^{1/2} & 0 &  0       & i\sqrt{1-Y} &0 \\
0 & 0  & 1& 0 & 0 & 0\\
0  & 0 &	0&  1 & 0 &	0 \\			
 0 &  i\sqrt{1-Y} &0 & 0    & Y^{1/2}& 0 \\
0 & 0 &   0 & 0 &0   & 1
\end{array}
\right), \quad 
\hat{S}_3=\left( 
\begin{array}{cccccc}
1& 0 &0 &0& 0  & 0\\
0 & Y^{1/2} & 0 & -i\sqrt{1-Y}    & 0 &0 \\
0 & 0  & 1 & 0 & 0 & 0\\
0  & -i\sqrt{1-Y}&	0&  Y^{1/2} & 0 &	0 \\			
 0 &  0 &0 & 0    & 1& 0 \\
0 & 0 &   0 & 0 &0   & 1
\end{array}
\right).
\label{do-b2-S51}
\end{equation} 
The last crossing point is between levels 1, 4 and 6, which is again described by the three-state bow-tie model: 
\begin{equation}
\hat{S}_4=\left( 
\begin{array}{cccccc}
X Y& 0 &0& i\sqrt{(1-X)(1+XY)} & 0  &  i\sqrt{X(1-Y)(1+XY)}\\
0 & 1 & 0 &  0         & 0 &0 \\
0 & 0  & 1 & 0 & 0 & 0\\
i\sqrt{(1-X)(1+XY)}   & 0&	0 &  X & 0 &	 -\sqrt{X(1-X)(1-Y)} \\			
 0 &  0&0   & 0    & 1& 0 \\
 i\sqrt{X(1-Y)(1+XY) }& 0 &    0 & -\sqrt{X(1-X)(1-Y)}  &0   & 1-X+X Y 
\end{array}
\right).
\label{do-b2-ham61}
\end{equation} 
The final scattering matrix of the model  (\ref{bos-distH4}) is given by 
\be
\hat{S}'=\hat{S}_4\hat{S}_3\hat{S}_2\hat{S}_1.
\label{full-skm}
\ee 
Let us introduce parameters 
$$
Q_X\equiv 1-X, \quad Q_Y\equiv 1-Y, \quad Q_{XY} \equiv 1+XY.
$$ 
Multiplying scattering matrices in (\ref{full-skm}) and taking the absolute values squared of all elements of $\hat{S}'$ we obtain the final matrix of transition probabilities:
\be
\hat{P}' =  \left( 
\begin{array}{cccccc}
X^2Y^2 & YQ_X Q_YQ_{XY} & Q_{XY}^2 Q_Y^2 & YQ_XQ_{XY} & 0 & XY^2Q_YQ_{XY}\\
0 &  Y^2 & YQ_XQ_YQ_{XY}& Q_Y &  X^2YQ_Y &XYQ_XQ_Y^2 \\
0 & 0 & X^2Y^2  &  0 & Q_XQ_{XY}& XQ_YQ_{XY} \\
Q_XQ_{XY}&     X^2Y Q_{XY}& 0 &  X^2Y & X^2Q_Y^2 & XQ_XQ_Y \\
0  &     Q_Y  & YQ_XQ_{XY}  & 0  & X^2Y  & XYQ_XQ_Y\\
XQ_YQ_{XY} &  XYQ_XQ_Y^2 & XY^2Q_YQ_{XY} & XYQ_{X}Q_Y & XQ_XQ_Y &(Q_X+XY^2)^2
\end{array}
\right).
\label{ptotS1}
\ee
Figures~\ref{add}(c-d) show perfect agreement of this prediction with results of our direct numerical simulations. This model example demonstrates basic nontrivial features of the semiclassical ansatz such as the possibility to describe simultaneous multiple level intersections and interference of semiclassical trajectories that connect  complex crossing points. 

\section{Model of Interacting Fermions}
\label{s-fermions}
Now, we turn to the question whether one can use our strategy in order to generate physically meaningful solvable MLZ models with combinatorial complexity. At least one such a model has been  already identified previously \cite{cQED-LZ}. It is important, however, to know how common such models are and how different they can be.
Here we will present quite a different from \cite{cQED-LZ} model, which describes nonlinear interactions of fermions.
\subsection{Six-state sector}

\begin{figure}
 \scalebox{0.35}[0.35]{\includegraphics{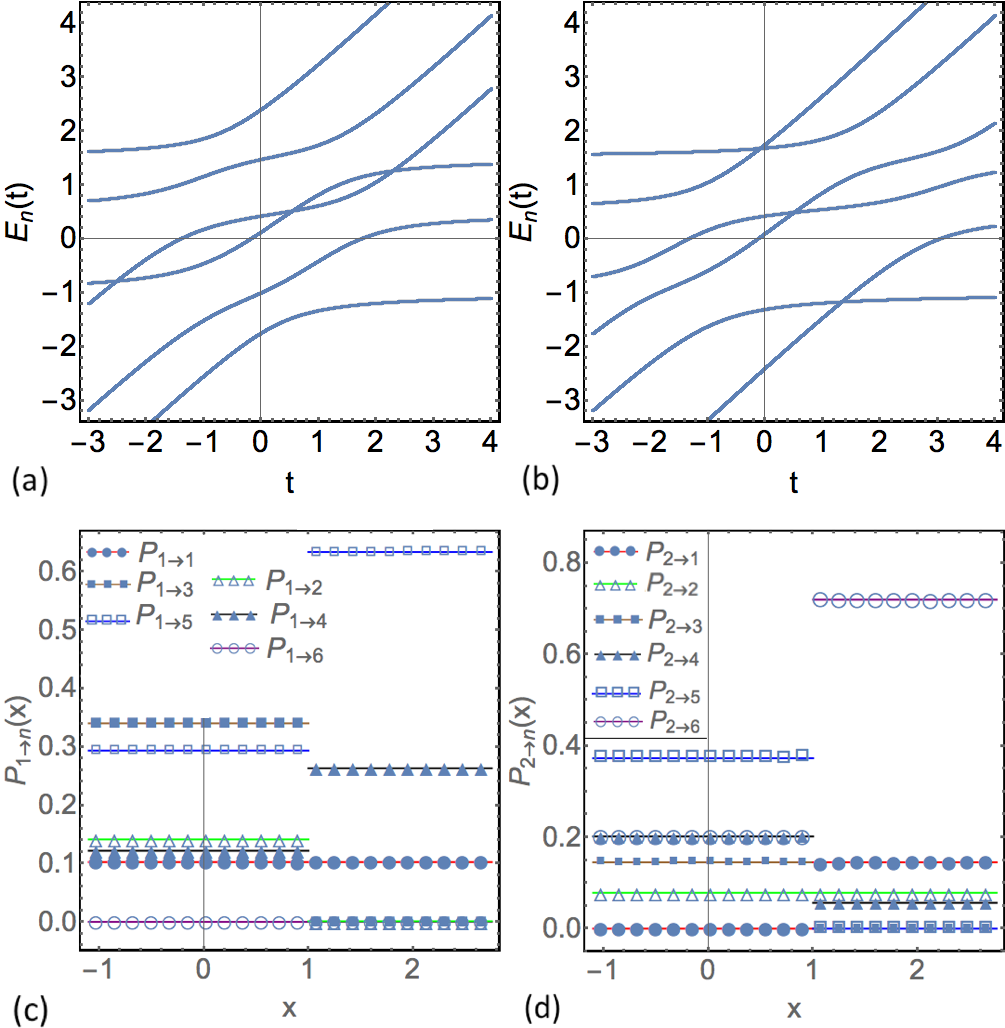}}
\hspace{-2mm}\vspace{-4mm}   
\caption{  (a-b) Eigenenergies of the Hamiltonian (\ref{six-ham}) as functions of $t$ at (a) $x=-0.5$ and (b) $x=2.5$.
Other parameters: $\beta=1$, $g_1=0.55$, $g_2=0.75g_1$, $g_3=0.85g_1$, $e_1=-1$, $e_2=0$, $e_3=1.5$. 
Despite large values of couplings, figures show three robust exact crossing points of energy levels.   (c-d) Numerical test of the  theoretical prediction for transition probabilities as function of coupling parameter $x$ in the six-state model with different initial conditions. (c) Transition probabilities from the 1st diabatic state to all diabatic states. Parameters: $e_1=0.45,\, e_2=0,\, e_3=-0.35$; $g_1=0.35,\, g_2=0.45,\, g_3=0.5$; $\beta=0.5$. (d) Transition probabilities from the 2nd diabatic state to all diabatic states. Parameters: $e_1=0.5,\, e_2=0.2,\, e_3=-0.2$; $g_1=g_2=g_3=0.45$; $\beta=0.5$. Evolution is simulated from $t=-2000$ to $t=2000$ with a time step $dt=0.00005$. Discrete points are numerical results and solid lines are theoretical predictions of Eqs~(\ref{six-p-in1})-(\ref{six-p-in2}).}
\label{six-fig}
\end{figure}

The logic that brought us to this model was the following.
We started with observation that there are two already known six-state solvable MLZ models with very similar diabatic level diagrams: 
\begin{equation}
\hat{H}_1=\left( 
\begin{array}{cccccc}
e_1+\beta t   & 0                   &0          		 & -g_2                  &-g_3    &0  		\\
0                  & e_2+\beta t  & 0           		& g_1                      &0 	&-g_3	 \\
0                 & 0         		& e_3+\beta t    		 &0                         & g_1	&g_2		 \\
-g_2		   & g_1		&0 					&e_1+e_2 & 0		& 0		\\			
-g _3          & 0                 & g_1   				& 0      		     & e_1+e_3 & 0	\\
0		 & -g_3		&g_2					&0				&0 	& e_2+e_3 
\end{array}
\right), \quad e_1<e_2<e_3, \quad \beta>0,
\label{six-ham-in1}
\end{equation}
and
\begin{equation}
\hat{H}_2=\left( 
\begin{array}{cccccc}
-e_1+\beta t   & 0                   &0          		 & -g_2                  &-g_3    &0  		\\
0                  & -e_2+\beta t  & 0           		& g_1                      &0 	&-g_3	 \\
0                 & 0         		& -e_3+\beta t    		 &0                         & g_1	&g_2		 \\
-g_2		   & g_1		&0 					&e_1+e_2	  & 0		& 0		\\			
-g _3          & 0                 & g_1   				& 0      		     & e_1+e_3 & 0	\\
0		 & -g_3		&g_2					&0				&0 	& e_2+e_3 
\end{array}
\right), \quad e_1<e_2<e_3, \quad \beta>0.
\label{six-ham-in2}
\end{equation} 
 The matrix of transition probabilities in  the model (\ref{six-ham-in1})  was derived in \cite{multiparticle} as a two-fermion extension of the DO model. The model   (\ref{six-ham-in2}) was found and solved only recently using ICs \cite{six-LZ}.   

Similarity between  Hamiltonians (\ref{six-ham-in1}) and (\ref{six-ham-in2}) suggests that they can be special instances of a more general model. The simplest candidate for such a generalization is the following
 6$\times$6 Hamiltonian matrix:
\begin{equation}
\hat{H}(t)=\left( 
\begin{array}{cccccc}
e_1(1-x)+\beta t   & 0                   &0          		 & -g_2                  &-g_3    &0  		\\
0                  & e_2(1-x)+\beta t  & 0           		& g_1                      &0 	&-g_3	 \\
0                 & 0         		& e_3(1-x)+\beta t    		 &0                         & g_1	&g_2		 \\
-g_2		   & g_1		&0 					&e_1+e_2	 & 0		& 0		\\			
-g _3          & 0                 & g_1   				& 0      		     & e_1+e_3 & 0	\\
0		 & -g_3		&g_2					&0				&0 	& e_2+e_3 
\end{array}
\right),
\label{six-ham}
\end{equation} 
where $x$ is a new parameter. 
At $x=0$ and $x=2$ the model (\ref{six-ham}) transfers into the models (\ref{six-ham-in1}) and  (\ref{six-ham-in2}),  respectively. 

It is straightforward to check that IC (i) is  always satisfied at $x\ne 1$, however, when $x$ crosses the value $x=1$ chronological order of  diabatic level crossigs changes. We tested validity of IC (ii) for both $x < 1$ and $x>1$ numerically and found that three crossing points of diabatic levels with zero direct coupling between them always resulted in exact crossings of adiabatic energy levels, as illustrated in Figs.~\ref{six-fig}(a-b). 

Since ICs are satisfied, we can apply the semiclassical ansatz to derive transition probability matrices. However, long calculations can be avoided because this analysis is completely analogous to the one in Ref.~\cite{six-LZ} that was used to solve the model (\ref{six-ham-in2}). Namely, in the case with $x>1$, topology of the diabatic level graph of  the model (\ref{six-ham}) and chronological order of level crossings is the same as for the graph of the model (\ref{six-ham-in2}), while coupling constants are also the same. So the transition probability matrix does not depend on $x$, as far as $x>1$, and solution derived in Ref.~\cite{six-LZ} for the model (\ref{six-ham-in2}) is equally applicable to any case with $x>1$. Similarly, at any $x<1$, topology of the graph of diabatic levels and chronological order of level crossings is the same as in the model (\ref{six-ham-in1}). Hence, the transition probability matrix that is given by the semiclassical ansatz  at any $x<1$ is the same as the one at $x=0$, which was found analytically in Ref.~\cite{multiparticle}.  

Let us introduce parameters:
\begin{equation}
p_{k}=e^{-2\pi g_k^2/\beta}, \quad q_{k}=1-p_k, \quad k=1,2,3.
\label{pppq}
\end{equation}
Final results for the two sectors of the model (\ref{six-ham}) read:
\begin{equation}
\hat{P}^{x<1}=\left( 
\begin{array}{cccccc}
p_2p_3   & q_1q_2p_3   &q_1q_3 & p_1q_2p_3                  &p_1q_3  &0  		\\
0 & p_1p_3     & p_1q_2q_3 &q_1p_3 	&q_1q_2q_3&p_2q_3	 \\
0 & 0  & p_1p_2 & 0	&q_1p_2&q_2		 \\
q_2 & q_1p_2	&0 	&p_1p_2  & 0		& 0		\\			
p_2q_3 & q_1q_2q_3  & q_1p_3  &p_1q_2q_3      		     & p_1p_3 & 0	\\
0 & p_1q_3 &p_1q_2p_3	&q_1q_3& q_1q_2p_3	& p_2p_3
\end{array}
\right),
\label{six-p-in1}
\end{equation}
and
\begin{equation}
\hat{P}^{x>1}=\left( 
\begin{array}{cccccc}
p_2p_3&0&0&q_2p_3&q_3&0\\
q_2q_1p_3&p_1p_3&0&p_2q_1p_3&0&q_3\\
q_1q_3&p_1q_3q_2&p_1p_2&0&p_3q_1p_2&p_3q_2\\
q_2p_1&q_1&0&p_1p_2&0&0\\
p_2q_3p_1&0&q_1&q_2q_3p_1&p_1p_3&0\\
0&p_1q_3p_2&p_1q_2&q_1q_3&p_3q_1q_2&p_2p_3
\end{array}
\right).
\label{six-p-in2}
\end{equation}

In Figs.~\ref{six-fig}(c-d) we show comparison of  theoretical predictions of the semiclassical ansatz, Eqs.~(\ref{six-p-in1})-(\ref{six-p-in2}),  with results of direct numerical simulations. Transition probabilities are plotted versus  parameter $x$. As predicted by the semiclassical ansatz, there is no dependence of such probabilities on $x$ except at the critical point  $x=1$, where  sharp change of behavior happens. Very small deviations from theoretical predictions can be noticed for some numerical points near $x=1$, however, this is explained by the fact that it takes  more time for transition probabilities to saturate in the vicinity of $x=1$ than the time interval considered in our numerical simulations.
We also note that parameters in Figs.~\ref{six-fig}(c-d) were intentionally chosen mainly from the nonperturbative domain $|e_i-e_j| \sim |g_{ij}|$  for $i<j=1,2,3$ in order to minimize the possibility that agreement with theory is accidental. 
\subsection{Multiparticle generalization} 

To search for further extension of the model (\ref{six-ham}) we  recall that the model (\ref{six-ham-in1}) was  obtained in Ref.~\cite{multiparticle} by populating the four-state DO model  with two noninteracting fermions. So,
the Hamiltonian of the model (\ref{six-ham-in1}) can be written as a quadradic Hamiltonian of four interacting fermionic modes \cite{multiparticle}. As the model (\ref{six-ham})  is different from (\ref{six-ham-in1}) only by constant terms on the main diagonal, we searched for the secondary quantized Hamiltonian of the model (\ref{six-ham}) by adding additional quartic terms to the fermionic representation of the model  (\ref{six-ham-in1}). We found the Hamiltonian 
\be
\hat{H}(t)= \beta t \hat{d}^{\dg} \hat{d} +\sum_{k=1}^{N-1}  \left[ e_k(1-x\hat{d}^{\dg} \hat{d} ) \hat{c}_k^{\dg} \hat{c}_k +g_k\left(\hat{d}^{\dg} \hat{c}_k +\hat{c}_k^{\dg} \hat{d} \right) \right],
\label{ham1}
\ee
where $\hat{c}_k$ and $\hat{d}$ are fermionic annihilation operators and $e_k$, $g_k$, $x$ are constants. The matrix form of  the Hamiltonian  (\ref{ham1}) is the same as in Eq.~(\ref{six-ham}) if we choose $N=4$ in the two-particle sector and the basis
\begin{eqnarray}
\nonumber |1\ra &\equiv& \hat{d}^{\dg} \hat{c}_1^{\dg} |0\ra, \quad |2\ra \equiv \hat{d}^{\dg} \hat{c}_2^{\dg} |0\ra, \quad |3\ra \equiv \hat{d}^{\dg} \hat{c}_3^{\dg} |0\ra,  \\
|4\ra &\equiv& \hat{c}_1^{\dg} \hat{c}_2^{\dg} |0\ra, \quad |5\ra \equiv \hat{c}_1^{\dg} \hat{c}_3^{\dg} |0\ra, \quad |6\ra \equiv \hat{c}_2^{\dg} \hat{c}_3^{\dg} |0\ra.
\label{basis-6}
\end{eqnarray}

\begin{figure}
 \scalebox{0.4}[0.4]{\includegraphics{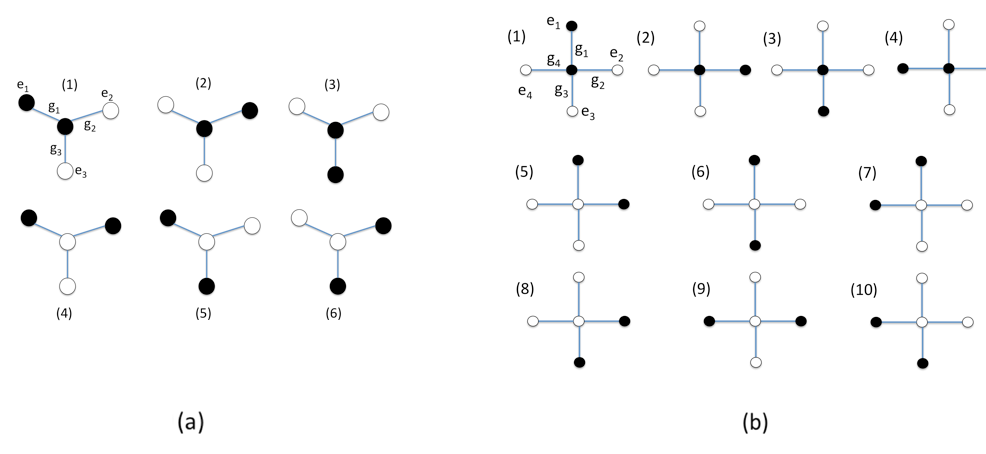}}
\hspace{-2mm}\vspace{-4mm}   
\caption{   (a) Six different states defined in (\ref{basis-6}) that can be produced by two fermions  residing in four different quantum dots. Filled/empty circles correspond to occupied/empty quantum dots; links correspond to allowed transitions between quantum dots when only one of those dots is occupied.   (b) Ten different states defined in (\ref{basis-10}) that can be produced by distributing two fermions among five different quantum dots. State numbering corresponds to our choice of diabatic state indexes. }
\label{six1-fig}
\end{figure}

The Hamiltonian (\ref{ham1}) conserves the number of fermions, $N_F$, which we can treat as a free integer parameter of the model.
We can also allow
$N$ to be an arbitrary positive integer. Generally, we can interpret $N$ as the number  of different ``quantum dots" in which spinless fermions of the model reside. There is a single energy level in each quantum dot so, due to the Pauli principle, maximum one fermion can reside in each dot. 
 One of such quantum dots, which is described by the operator $\hat{d}$, is special: fermions can produce transitions between different states only via this dot, which is coupled to all other dots by
tunneling amplitudes $g_k$, $k=1,\ldots, N-1$, as shown in Fig.~\ref{six1-fig}(a) for the case $N=4$.  Moreover, we assume that  only the energy of the fermion in this dot is time-dependent, e.g., due to a linearly time-dependent gate voltage. 
Parameter $x$ describes the nonlinear effect of coupling of the special dot to other dots. This coupling is activated every time the special dot is occupied. Physical meaning of parameter $e_k$ is  merely the energy of a fermion when it resides in the quantum  dot with index $k \in 1,\ldots, N-1$. 

Does the Hamiltonian (\ref{ham1}) satisfy ICs? In this article, we do not provide the definite answer but we do find strong evidence in support of the conjecture that this model is, indeed, solvable. For $N_F>1$, 
 there are $n=N!/[N_F! (N-N_F)!]$ different states in which one can place $N_F$ fermions among $N$ sites. For example, at $N=4$ and $N_F=2$, there are six states available for dynamics, and we already provided the evidence that this model satisfies ICs and its  solution coincides with prediction of the semiclassical ansatz.

 Let us next consider the case $N_F=1$, and look at evolution in the basis of Fock states: $|k\ra \equiv \hat{c}_k^{\dg} |0\ra$, $|N \ra \equiv \hat{d}^{\dg} |0\ra$, where $|0\ra$ is the vacuum state of fermions. The Hamiltonian in this basis has the form
\begin{equation}
\hat{H}^{(N,1)}=\left( 
\begin{array}{ccccc}
e_1 & 0                   &\ldots     & 0                  &g_1  \\
0                  & e_2  & 0           & \ldots           &g_2  \\
\vdots          & \vdots          & \ddots  &\vdots           & \vdots \\
0		   & 0 		& \ldots 	&e_{N-1}		     & g_{N-1}	\\			
g _1                  & g_2                  & \ldots   & g_{N-1}       &\beta t
\end{array}
\right),
\label{do-ham}
\end{equation} 
where the upper index in $\hat{H}^{(N,1)}$ means that the Hamiltonian (\ref{ham1}) is restricted to one fermion that can be in $N$ different states. 
The Hamiltonian (\ref{do-ham})  is precisely the $N$-state  DO model, which is exactly solvable \cite{do} and for which ICs are trivially satisfied. 
 
\begin{figure}
\scalebox{0.4}[0.4]{\includegraphics{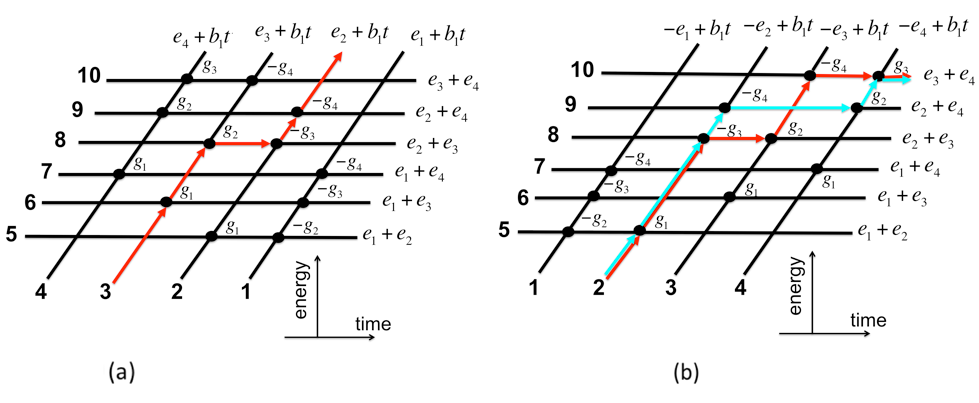}}
\hspace{-2mm}\vspace{-4mm}   
\caption{  (a) Diabatic levels of the ten-state sector of the model (\ref{ham1}) at $x=0$. Red color arrows show the unique semiclassical path that connects levels 3 and 2. (b) Diabatic levels of a ten-state model at $x=2$. Red and blue color arrows illustrate interfering semiclassical paths that connect levels 2 and 10. }
\label{ten2-fig1}
\end{figure}

\subsection{Ten-state sector}
\begin{figure}
\scalebox{0.35}[0.35]{\includegraphics{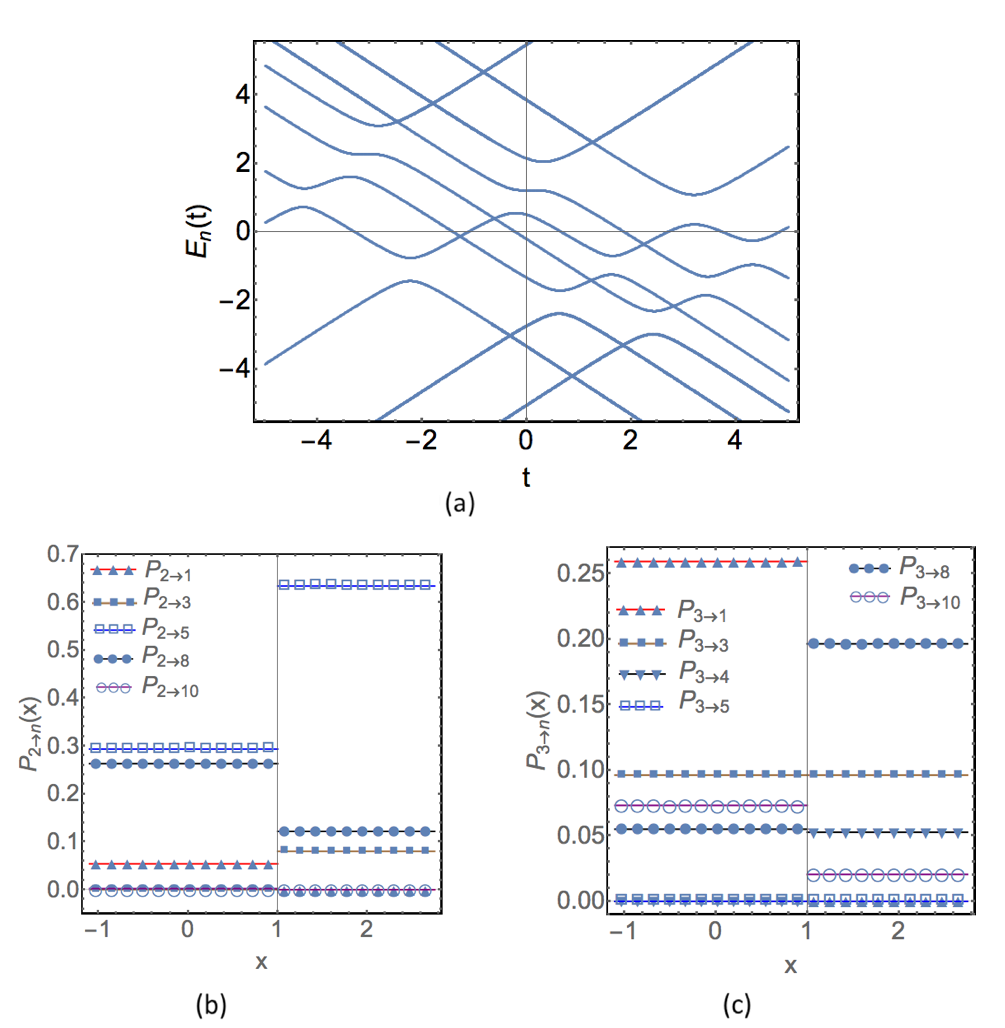}}
\hspace{-2mm}\vspace{-4mm}   
\caption{(a) Eigenenergies of the Hamiltonian (\ref{ham1})  at $N=5$, $N_F=3$ as functions of $t$. For convenience, a term   $-\hat{1}\beta t /2$ was added to the Hamiltonian, where $\hat{1}$ is a unit matrix; in our ten-state model it merely equally changes slopes of all levels. Parameters: $x=2.5$, $\beta=1$, $g_1=0.44$, $g_2=0.36$, $g_3=0.28$, $g_4=0.34$, $e_1=2.5$, $e_2=1.3$, $e_3=-0.6$, $e_4=-2.7$.
Despite large values of couplings, figure shows twelve exact crossing points of energy levels. (b-c) Numerical test of semiclassical predictions for transition probabilities as function of the quartic coupling parameter $x$ for initially occupied (b) level-2, (c) level 3. Parameters:  $g_1=0.4$, $g_2=0.35$,
$g_3=0.45$, $g_4=0.3$, $e_1=-0.75$, $e_2=-0.25$, $e_3=0.4$, $e_4=1$, $\beta=1$.} 
\label{ten1-fig}
\end{figure}

The next in complexity sector of the model (\ref{ham1}) has two fermions that can occupy any of five quantum dots, as shown in Fig.~\ref{six1-fig}(b). In the basis
\begin{eqnarray}
\label{basis-10}
 |1\ra &\equiv& \hat{d}^{\dg} \hat{c}_1^{\dg}  |0\ra, \quad |2\ra \equiv \hat{d}^{\dg} \hat{c}_2^{\dg} |0\ra, \quad
|3\ra \equiv \hat{d}^{\dg} \hat{c}_3^{\dg}  |0\ra, \quad |4\ra \equiv \hat{d}^{\dg} \hat{c}_4^{\dg} |0\ra, \quad
|5\ra \equiv \hat{c}_1^{\dg} \hat{c}_2^{\dg}  |0\ra, \\
\nonumber |6\ra &\equiv& \hat{c}_1^{\dg} \hat{c}_3^{\dg} |0\ra, \quad
 |7\ra \equiv \hat{c}_1^{\dg} \hat{c}_4^{\dg}  |0\ra, \quad |8\ra \equiv \hat{c}_2^{\dg} \hat{c}_3^{\dg} |0\ra, \quad
 |9\ra \equiv \hat{c}_2^{\dg} \hat{c}_4^{\dg}  |0\ra, \quad |10\ra \equiv \hat{c}_3^{\dg} \hat{c}_4^{\dg} |0\ra,
\end{eqnarray}
the Hamiltonian (\ref{ham1}) is a 10$\times$10 matrix, which we will not show here explicitly. Diabatic levels of this model are shown in Fig.~\ref{ten2-fig1}. The first IC is trivial  but tedious to verify, so we will skip this step here.  In Fig.~\ref{ten1-fig}(a), we show   adiabatic energies of this Hamiltonian that we calculated numerically at $x=2.5$. One can visually see that there are 
twelve exact crossing points in this figure, as it is needed to satisfy IC (ii).

Here we note that not all exact crossings in Fig.~\ref{ten1-fig}(a) are robust in the sense that by increasing couplings some of the crossings can annihilate with each other. However, as it was discussed in \cite{cQED-LZ}, IC (ii) requires   presence of a specific number of exact crossing points only at finite but sufficiently small values of all couplings.


\subsubsection{Transition probabilities at $x<1$}


If we restrict attention only to  values $x<1$, topology of the graph of diabatic states is the same as at $x=0$; hence we can safely set $x=0$ in order to determine the transition probability matrix.
The point $x=0$ corresponds to the noninteracting case with a quadratic fermionic Hamiltonian. In the Heisenberg's picture, evolution of operators  has then the same form as evolution of amplitudes in the DO model \cite{multiparticle}. Consider  the DO model with $N$ levels marked so that the level with index $N$ has the positive slope and levels with indexes $i=1, \ldots, N-1$ belong to the band of parallel levels with diabatic energies $e_1<e_2< \ldots <e_{N-1}$. 
Solution of the Heisenberg equation for annihilation operators  can then be written in the form 
\be
\hat{c}_k (+\infty) = \sum_{l=1}^N S_{kl} \hat{c}_l (-\infty),
\label{heis}
\ee
where we identify $\hat{d}\equiv \hat{c}_N$. Transition amplitudes in the DO model are known \cite{be,do,multiparticle}: 
\begin{eqnarray}
\label{scattdo0}
S_{NN}&=&\left( \prod_{k=1}^{N-1} p_k \right)^{1/2}, \\
S_{nN} &=& i \left( q_n  \prod_{k=1}^{n-1} p_k \right)^{1/2},\\
S_{Nn} &=& i \left( q_n  \prod_{k=n+1}^{N-1} p_k \right)^{1/2}, \\
S_{nm}^{(n>m)} &=& - \left( q_nq_m  \prod_{k=m+1}^{n-1} p_k \right)^{1/2}, \\
S_{nm}^{(n<m)} &=& 0,
\label{scattdo1}
\end{eqnarray}
where we dropped the coupling-dependent phases of  amplitudes except the imaginary unit factors that are acquired at every turning to another level. It was discussed in \cite{four-LZ} that all other phase factors cancel in semiclassical calculations of transition probabilities in all known exactly solvable models, and we assume this is happening here too. We do not pursue the mathematical proof of this fact because we check the final result numerically.

 Consider a sector of such a model with $M$ fermions. Let indexes $\gamma_1, \ldots, \gamma_M$ correspond to quantum dots that are initially populated with fermions at $t\rightarrow -\infty$, and let $\alpha_1, \ldots, \alpha_M$ be indexes of the dots populated with fermions at $t\rightarrow +\infty$ ($\alpha_1<\alpha_2<\ldots < \alpha_M$ etc.).
The transition probability between such two states is the same as the product of average populations of final dots: 
\begin{widetext}
\be
P_{\gamma_1, \ldots, \gamma_M \rightarrow \alpha_1, \ldots, \alpha_M} = \la \gamma_1, \ldots, \gamma_M | \hat{c}_{\alpha_1}^{\dg}(+\infty)  \hat{c}_{\alpha_1} (+\infty) \ldots \hat{c}_{\alpha_M}^{\dg}(+\infty)  \hat{c}_{\alpha_M} (+\infty)
 | \gamma_1, \ldots, \gamma_M \ra,
\label{pr1}
\ee
\end{widetext} 
where
\be
 | \gamma_1, \ldots, \gamma_M \ra \equiv \hat{a}_{\gamma_1}^{\dg} (-\infty) \ldots \hat{a}_{\gamma_M}^{\dg} (-\infty) |0\ra,
\label{pr2}
\ee
and relation between combined indexes, such as $\gamma_1, \ldots, \gamma_M$, and numbering of diabatic levels is explained in Fig.~\ref{six1-fig} for sectors with six and ten interacting states.
Simple algebra leads to the final result: 
\be
P_{\gamma_1, \ldots, \gamma_M \rightarrow \alpha_1, \ldots, \alpha_M} = |{\rm Det} (\hat{Q}) |^2,
\label{pr3}
\ee
where 
\be
\hat{Q} =\left( 
\begin{array}{cccc}
S_{\alpha_1\gamma _1} & S_{\alpha_1 \gamma_2} &\ldots & S_{\alpha_1 \gamma_M} \\
S_{\alpha_2 \gamma_1} & S_{\alpha_2\gamma_2} & \ldots & \vdots \\
\vdots &  \vdots & \ddots &\vdots \\
S_{\alpha_M \gamma_1}& \cdots & \cdots& S_{\alpha_M \gamma_M}
\end{array}
\right).
\label{pr4}
\ee
Equations~(\ref{pr3})-(\ref{pr4}) with (\ref{scattdo0})-(\ref{scattdo1}) provide a simple formal algebraic solution of the problem at $x=0$ and consequently at any $x<1$.

 Physically more useful characteristic can be the average population of the site  (quantum dot) $\alpha$ irrespectively of occupations of other states. For independent fermions,  this characteristic was previously obtained in \cite{multiparticle}: 
\be
P_{\alpha} =\la \gamma_1, \ldots, \gamma_M | \hat{c}_{\alpha}^{\dg}(+\infty)  \hat{c}_{\alpha} (+\infty)| \gamma_1, \ldots, \gamma_M \ra = \sum_{j=1}^{M} |S_{\alpha \gamma_j}|^2.
\label{pr5}
\ee
So, we can only add here that Eq.~(\ref{pr5}) is equally valid for the case with quartic interactions at arbitrary $x<1$.

Let us now restrict to the case $N=5$ and $N_F=2$, which is illustrated in Fig.~\ref{ten2-fig1}(a). The  initially occupied state 3 in Fig.~\ref{ten2-fig1}(a) corresponds to initially occupied sites $\gamma_1=3$ and $\gamma_2=5$ in Fig.~\ref{six1-fig}(b). Finally occupied level 2  in Fig.~\ref{ten2-fig1}(a), for example,  corresponds to finally occupied quantum dots $\alpha_1=2$ and $\alpha_2=5$ in Fig.~\ref{six1-fig}(b).

 For  initial state 3 (see Fig.~\ref{six1-fig}(b)) and arbitrary final state $n=1,\ldots,10$, Eq.~(\ref{pr3}) then predicts the following vector of transition probabilities:  
\be
P_{3\rightarrow n}^{x<1} = \left(
q_1q_3p_4,\, p_1q_2q_3p_4,\, p_1p_2p_4,\, 0,\, 0,\, q_1p_3,\, q_1q_3q_4,\, p_1q_2p_3,\, p_1q_2q_3q_4,\, p_1p_2q_4
\right).
\label{pr6}
\ee
Similarly, if initial level is 2 then transition probability vector is 
\be
P_{2\rightarrow n}^{x<1} = \left(
q_1q_2p_3p_4,\, p_1p_3p_4,\ 0,\, 0,\, q_1p_2,\, q_1q_2q_3,\, q_1q_2p_3q_4,\, p_1q_3,\, p_1p_3q_4,\, 0\right).
\label{pr66}
\ee

\subsubsection{Transition probabilities for $x>1$ in ten-state sector}
We do not have a simple formula for $x>1$. Behavior of transition probabilities in this case is more complex because of the possibility of interference between different semiclassical trajectories. For example, there are two paths that connect initial level 2 and final level 10 in Fig.~\ref{ten2-fig1}(b), which we mark by red and blue arrows. 
So, we have to apply the semiclassical ansatz, which is equivalent to deriving all semiclassical paths that connect pairs of initial and final states in the diabatic level diagram and summing amplitudes of all these paths assuming that contributions from all level crossings to each path amplitude factorize. 

According to the semiclassical rules, each time a trajectory turns to another level at a crossing point with some pairwise coupling $g_k$
 it receives the factor $\pm i \sqrt{q_k}$, where ($\pm$) is the sign of the coupling. The amplitude of the trajectory marked by red arrows  in Fig.~\ref{ten2-fig1}(b) is then
$$
S_{\rm red} = (-i) i (-i) \sqrt{p_1q_3q_2q_4p_3}.
$$
Similarly, the trajectory marked by blue arrows has the amplitude
$$
S_{\rm blue} = -i^3 \sqrt{p_1p_3q_4q_2q_3}.
$$
Comparing them we find that amplitudes of red and blue trajectories have the same absolute magnitude but different signs. Therefore, 
the total amplitude is $S_{10,2} = S_{\rm red} +S_{\rm blue}  =0$. Similarly, we can find other transition probabilities starting with initial level 2:
\be
P_{2\rightarrow n}^{x>1} = \left(
0,\, p_1p_3p_4,\, p_1q_3q_2p_4,\, p_1q_2q_4,\, q_1,\, 0, \, 0,\, p_1q_3p_2,\, p_1p_3q_4p_2,\, 0
\right).
\label{pr7}
\ee
If initial state is 3 then there is no path interference and transition probability vector is 
\be
P_{3\rightarrow n}^{x>1} = \left(
0,\, 0,\, p_1p_2p_4,\, p_1p_2q_4q_3,\, 0,\, q_1,\, 0,\, p_1q_2,\, 0,\, p_1p_2q_4p_3
\right).
\label{pr77}
\ee

We compare predictions of Eqs.~(\ref{pr6})-(\ref{pr77}) with nuremical results in Fig.~\ref{ten1-fig}(b-c) and again find perfect agreement. Thus, we verified that ICs are valid for, at least, three simplest sectors of the model (\ref{ham1}), and that numerically calculated transition probabilities in these sectors coincide with analytical predictions of the semiclassical ansatz within three significant digit precision for all considered parameter values. Complexity of the phase space of the model (\ref{ham1}) is growing very quickly with increasing numbers of fermions and quantum dots, which makes further numerical tests problematic but already available results do support our conjecture that the full model (\ref{ham1}) is integrable.

\section{Discussion}

We demonstrated the strategy to search for new solvable MLZ models. The idea is to start with a graph of some already known solvable model and then distort it keeping only topology of the original graph and allowing a broader choice of parameters. 
ICs  can then be converted to a set of algebraic equations on new parameters. Thus, IC (i) of zero area inside any closed path of a graph leads to constraints on slopes of diabatic levels. Treating couplings as small, one can then derive the necessary conditions that such couplings should satisfy in order to agree with IC (ii), which is achieved with the quantum perturbation theory. Sufficiency of these conditions is tested numerically. Following this path, we constructed several new solvable models of different complexity, and we provided detailed numerical check of our predictions for transition probabilities in these models. 

Based on our findings, it seems that solvable MLZ models are much more common than it has been previously expected. By creating symmetrized products of already solved systems or populating these systems with some number of noninteracting bosons and fermions, as it is described in \cite{multiparticle}, we can generate  more complex graphs that, in turn, can be distorted, leading to new integrable MLZ models and so on. Moreover, special cases when some of the diabatic levels become parallel produce additional opportunities to distort parameters without breaking ICs. Such opportunities to find new solvable models using distortions  can be combined also with previously developed methods to derive systems with smaller phase space sizes from the bigger models.  This can be done either by identifying the bigger model as a composite one \cite{constraints} or by considering limits with level degeneracies and decoupling some of the states  in these limits \cite{gbow-tie,cQED-LZ}.

While the origin of the MLZ integrability remains puzzling, discoveries of new solvable models can have important consequences. For example, we found that  transition probability matrices in newly solved models had the same symmetries as in models that had been rigorously solved by other means.  Understanding such relations may lead to the rigorous proof of ICs  and validity of the semiclassical ansatz. We also find it surprising that, in all promising candidate models, perturbatively derived constraints on coupling constants have lead to truly exact eigenvalue crossings. Formally, perturbatively derived constraints are only necessary conditions that do not have to be sufficient. It seems that, somehow, imposing IC (i) facilitates the emergence of exact crossing points so that a simple perturbative test, which is performed up to some finite order of perturbation theory, provides not only necessary but also sufficient conditions for existence of such points. This observation is one of already many  surprisingly fortunate but puzzling properties of MLZ models.

Exact and approximate methods to connect asymptotic behavior of ordinary differential equations are of general interest to mathematics and mathematical physics \cite{joye-LZ,aoki-LZ}. For example, the   most physically important special functions are solutions of differential equations with well understood asymptotics. Future studies should answer the question whether solvable MLZ models represent a new class of physically important solvable differential equations or such models can be understood in terms of the already well studied special functions, such as the generalized hypergeometric  function.

Finally, some of the solutions that have been obtained with ICs are  important for practical reasons. For example, our fermionic model in Sec.~\ref{s-fermions} describes the process of a quantum level interaction with a fermionic bath when this level is driven by a time-dependent gate voltage. Such models have been of interest previously  due to the possibility for control of quantum states with minimal dissipation \cite{multiparticle,levitov-lz1}. So, with MLZ solutions, it becomes possible to explore complex explicitly driven quantum mechanical systems analytically and exactly. We hope that many other physically interesting processes will be eventually understood with such solvable  models.

\appendix
\section{Counterexample to previous version of the first IC}
\label{a-counter}
\begin{figure}
\scalebox{0.35}[0.35]{\includegraphics{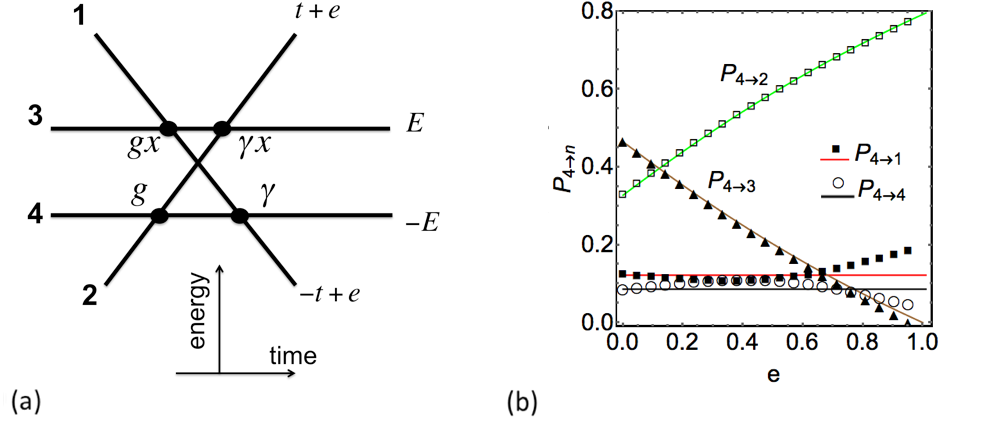}}
\hspace{-2mm}\vspace{-4mm}   
\caption{(a) Diabatic level diagram of the model that breaks condition (i) as defined in this article but satisfies ICs defined in \cite{six-LZ}. 
(b) Numerical test showing that  the semiclassical ansatz (solid curves) does not fit results of numerical simulations (discrete points). Parameters: $E=1$, $g=0.5$, $\gamma=0.75g$, and $x$ is given by Eq.~(\ref{xeE}). Visible agreement for $P_{4\rar 2} $ is not precise either: systematic deviations from the semiclassical ansatz were found in the third significant digit (not shown).  }
\label{four-fig2}
\end{figure}

Our definition of the first integrability condition in Sec.~\ref{sec-IC} is different from the original definition in \cite{six-LZ}. Previously, condition (i) requested that if, on the diabatic level diagram, there are more than one semiclassical paths that connect one initial level at $t=-\infty$ with another level at $t=+\infty$ then dynamic phases, i.e. integrals of diabatic energies over time, of such trajectories should be the same. The latter property is actually the consequence of IC (i)  defined in the present article because the dynamic phase is the area that a trajectory in the level diagram sweeps over the time axis. Equality of such phases  means that the path that goes forward along one of these trajectories  and then returns to the initial point along the other trajectory encloses zero area  in this diagram.

The new definition, however, is more restrictive because it includes the case when different trajectories do not interfere in the physical sense but the graph still has closed loops. We show an example of such a model  in Fig.~\ref{four-fig2}(a). One can verify numerically that IC (ii) is satisfied when
\begin{equation}
x=\sqrt{(E-e)/(E+e)},
\label{xeE}
\end{equation}
i.e., at such relations among parameters there is an exact crossing point of adiabatic levels near the intersection of levels 1 and 2.  As for the first IC (i), it is not satisfied at $e\ne 0$ because the graph has a closed path that encloses a finite area. So, semiclassical ansatz should fail. 

On the other hand, each semiclassically allowed trajectory in Fig.~\ref{four-fig2}(a) is unique in the sense that there is no other path that respects causality and connects the same pair of levels, i.e., there are no physically interfering trajectories.  So, if definition in Ref.~\cite{six-LZ} is generally true then the semiclassical ansatz should apply.

Figure~\ref{four-fig2}(b) compares prediction of the semiclassical ansatz for this model with results of numerical simulations. Although the semiclassical ansatz gives a reasonable fit to the data, there is noticable disagreement. This makes us to conclude that the more restrictive definition of condition (i) in Sec.~\ref{sec-IC} of the present article  should be preferred. So far, we are not aware of any MLZ model that would satisfy new  version of ICs (i)-(ii) but would not be solvable by the semiclassical ansatz.

\section{Analogy between first integrability condition and WKB description of classically integrable systems}
\label{sec:MLZ-graph}

Here we would like to point that
the geometric interpretation of the first integrability condition, presented in Sec.~\ref{sec-IC}, may provide a link between the integrable MLZ problems and semiclassical WKB description of classically integrable systems.
Consider the graph $G = (G_{0}, G_{1})$ associated with the diabatic level diagram of a given $n \times n$ MLZ problem.
The vertices in $ G_{0}$ of this graph are given by those intersection points for which the coupling between the intersecting diabatic levels is non-zero, and the edges in $ G_{1}$ represent the segments of the straight lines between two consecutive scattering events. Let $X$ be the geometric realization of the graph $G$, i.e., 
 $X$ stands for the graph that is viewed as a set of segments $s_{e}$, associated with its edges $e \in G_{1}$, connected to each other at the vertices, rather than just a collection $G = (G_{0}, G_{1})$ of edges and vertices.
By definition, every point of $X$ defines diabatic energy $\omega(t)$ of a diabatic state represented by this point.  Hence, the geometric space $X$ of our MLZ problem has a natural immersion $f : X \looparrowright \mathbb{R}^{2}$, which is what is shown in the diabatic level diagram. 

In the $(t, \omega)$ plane we have a rank $1$ form $\alpha = \omega dt$, which, being restricted (strictly speaking pulled-back) to $X$ gives rise to a form $f^{*}\alpha$ on $X$. 
The first integrability condition (area rule) is then equivalent to the requirement that for any cycle $C$ in $X$ we have
\begin{eqnarray}
\label{inegrability-geom} \int_{C}f^{*}\alpha = 0,
\end{eqnarray}
which, due to the quasi-one-dimensional nature of $X$ is equivalent to existence of a global function ${\cal S}: X \rightarrow \mathbb{R}$ with $d{\cal S} = f^{*}\alpha$, or, stated differently,  ${\cal S} = \int f^{*}\alpha$. We will refer to ${\cal S}$ as the action function.

 For $1$-dimensional quantum systems, the WKB approximation is based on the action function that can be represented as ${\cal S} = \int \alpha$, with $\alpha = p dx$. The WKB form $\psi(x) \sim \exp\left(i\hbar^{-1}{\cal S}(x)\right)$ of an eigenstate may not be  generalizable to the multi-dimensional case, however if an $n$-dimensional system is integrable, i.e, it has $n$ mutually commuting integrals $I_{j}$ of motion, a multidimensional counterpart of a WKB wave function can be obtained. Indeed, switching to the classical limit and further fixing the values of all integrals of motion $I_{j}(p, x) = E_{j}$, where $E_j$ are some constants for $j = 1, \ldots, n$, we can solve these equations with respect to $p$, resulting in the set of functions $p_{j}(x)$, which give rise to a $1$-form $\alpha = \sum_{j}p_{j} dx^{j}$. It is possible to show that the integrability condition $\{I_{j}, I_{k}\} = 0$, where $\{\ldots \}$ stands for Poisson brackets for all $j$ and $k$, is equivalent to the closeness condition $d\alpha = 0$, the latter meaning that one can define a global action function ${\cal S} = \int \alpha$ that determines the zero-order approximation for a semiclassical WKB function that has exactly the same form as in $1$-dimensional case. 
 
 Summarizing the analogy: in both cases, namely MLZ problems and quantized integrable systems, the integrability condition, at least in the lowest (principle) order of the semiclassical expansion, boils down to the existence of a globally defined action function ${\cal S}$. We would like to admit explicitly that at this point we do not see how exactly the above analogy can be applied. It may mean, for example, that integrable MLZ systems describe some ballistic, i.e.  large kinetic energy, limits of  integrable  many-body $1$-dimensional quantum systems. Finally, we note that this analogy resembles the property, found at least in some MLZ models, that solvable MLZ Hamiltonians have nontrivial commuting operators 
 that depend polynomially on time \cite{yuzbashyan-LZ}.

{\it Acknowledgment}. 
This work
was carried out under the auspices of the National Nuclear
Security Administration of the U.S. Department of Energy at Los
Alamos National Laboratory under Contract No. DE-AC52-06NA25396. V.Y.C. was supported by the National Science Foundation under Grant No. CHE-1111350. N.A.S. also thanks the support from the LDRD program at LANL.

\end{document}